\documentclass{article}
\pdfoutput=1
\usepackage{graphicx}  
\usepackage{amsmath}   
\usepackage[compress]{cite}
\usepackage{amssymb}   
\usepackage{bm} 
\usepackage{dcolumn}
\usepackage{color}
\usepackage{mathrsfs}
\usepackage{amsfonts}
\usepackage{varioref}
\usepackage{mathrsfs}
\usepackage{graphicx}
\usepackage{latexsym}
\usepackage{amsmath}
\usepackage{amssymb}
\usepackage{textcomp}
\usepackage{amsbsy}
\usepackage{graphics}
\usepackage{epstopdf}
\usepackage{color}
\usepackage[caption=false]{subfig}
\RequirePackage[colorlinks,citecolor=blue,urlcolor=magenta,linkcolor=blue]{hyperref}
\input epsf

\labelformat{section}{Section #1} 
\labelformat{subsection}{Section #1} 
\labelformat{subsubsection}{Section #1}
\labelformat{subsubsubsection}{Section #1}
\labelformat{equation}{Eq.~(#1)} 
\labelformat{figure}{Fig.~#1} 
\labelformat{subfigure}{Fig.~\thefigure#1} 
\labelformat{table}{Tab.~#1} 
\labelformat{appendix}{Appendix #1}
\title{Bouncing cosmology in a curved braneworld}

\author{Indrani Banerjee$^{1}$\footnote{banerjeein@nitrkl.ac.in}~,
Tanmoy Paul$^{2}$\footnote{pul.tnmy9@gmail.com}~
and Soumitra SenGupta$^{3}$\footnote{tpssg@iacs.res.in} \\
\small{$^{1}$Department of Physics and Astronomy, National Institute of Technology, Rourkela-769008, India }\\
\small{$^{2}$ Department of Physics, Chandernagore College, Hooghly - 712 136, India}\\
\small{$^{3}$ School of Physical Sciences, Indian Association for the Cultivation of Science, Kolkata-700032, India}}

\date{ }  
\begin{document}
  
\maketitle 
\begin{abstract}
We explore the possibility of a non-singular bounce in our universe 
from a warped braneworld scenario with dynamical branes and a non-zero brane cosmological constant. Such models naturally incorporate a 
scalar sector known as the radion originating from the modulus of the theory. The presence of brane cosmological constant renders the 
branes to be non-flat and gives rise to a potential and a non-canonical kinetic term for the radion field in the four dimensional effective action. 
The kinetic term exhibits a phantom-like behavior within the domain of evolution of the modulus which leads to a violation of the null-energy 
condition often observed in a bouncing universe. The interplay of the radion potential and kinetic term enables the evolution of the radion 
field from a normal to a phantom regime where the universe transits from a contracting era to an expanding epoch through a non-singular bounce. 
Analysis of the scalar and tensor perturbations over such background evolution 
reveal that the primordial observables e.g., the amplitude 
of scalar perturbations $\mathcal{A}_s$, tensor to scalar ratio $r$ and the scalar spectral index $n_s$ are in agreement with 
the current constraints reported by the Planck satellite. The implications are discussed.

\end{abstract}
\section{Introduction}
\label{Intro}
One of the major challenges in modern theoretical cosmology is to explain the early stage of the universe, in particular, whether the universe 
emerged from an initial singularity (also known as the Big-Bang singularity) or the universe underwent a non-singular bounce 
leading to a possible singularity free expansion of the universe.
Some of the early universe scenarios, proposed so far, that can generate an almost scale invariant power spectrum and hence 
confront the observational constraints 
are the \emph{inflationary scenario} \cite{guth,Linde:2005ht,Langlois:2004de,Riotto:2002yw,barrow1,barrow2,nb1,Baumann:2009ds}, 
the \emph{bouncing universe} \cite{Brandenberger:2012zb,Brandenberger:2016vhg,Battefeld:2014uga,Novello:2008ra,Cai:2014bea,Cai:2016thi,Cai:2017tku,
deHaro:2015wda,Lehners:2011kr,Lehners:2008vx,
Cheung:2016wik,Cai:2016hea,Cattoen:2005dx,Li:2014era,Brizuela:2009nk,Cai:2013kja,Quintin:2014oea,Cai:2013vm,Poplawski:2011jz,
Koehn:2015vvy,Odintsov:2015zza,Nojiri:2016ygo,Odintsov:2015ynk,Odintsov:2020zct,Koehn:2013upa,Battarra:2014kga,Martin:2001ue,Khoury:2001wf,
Buchbinder:2007ad,Brown:2004cs,Hackworth:2004xb,Nojiri:2006ww,Johnson:2011aa,Peter:2002cn,Gasperini:2003pb,Creminelli:2004jg,Lehners:2015mra,
Mielczarek:2010ga,Lehners:2013cka,Cai:2014xxa,Cai:2007qw,Cai:2010zma,Avelino:2012ue,Barrow:2004ad,Haro:2015zda,Elizalde:2014uba,Das:2017jrl,
Cai:2008qw,Finelli:2001sr,Cai:2011ci,Haro:2015zta,Cai:2011zx,Haro:2014wha,Brandenberger:2009yt,deHaro:2014kxa,Odintsov:2014gea,
Qiu:2010ch,Bamba:2012ka,deHaro:2012xj,Nojiri:2019lqw,Elizalde:2020zcb,WilsonEwing:2012pu}, 
the \emph{emergent universe scenario} \cite{Ellis:2003qz,Paul:2020bje,Li:2019laq,Cai:2012yf,Dutta:2016ual,Bag:2014tta} 
and the \emph{string gas cosmology} \cite{Brandenberger:2008nx,Brandenberger:2011et,
Brandenberger:1988aj,Kripfganz:1987rh,Battefeld:2005av,Nayeri:2005ck,Brandenberger:2006vv}.

In this work, we study the bouncing scenario in a non-flat warped braneworld model. The bouncing scenario 
consists of two eras$-$an era of contraction and an era of expansion of the scale factor, both the eras being connected by a non-singular bounce 
\cite{Brandenberger:2012zb,Brandenberger:2016vhg,Battefeld:2014uga,Novello:2008ra,Cai:2014bea,Cai:2016thi,Cai:2017tku,
deHaro:2015wda,Lehners:2011kr,Lehners:2008vx,
Cheung:2016wik,Cai:2016hea,Cattoen:2005dx,Li:2014era,Brizuela:2009nk,Cai:2013kja,Quintin:2014oea,Cai:2013vm,Poplawski:2011jz,
Koehn:2015vvy,Odintsov:2015zza,Nojiri:2016ygo,Odintsov:2015ynk,Odintsov:2020zct,Koehn:2013upa,Battarra:2014kga,Martin:2001ue,Khoury:2001wf,
Buchbinder:2007ad,Brown:2004cs,Hackworth:2004xb,Nojiri:2006ww,Johnson:2011aa,Peter:2002cn,Gasperini:2003pb,Creminelli:2004jg,Lehners:2015mra,
Mielczarek:2010ga,Lehners:2013cka,Cai:2014xxa,Cai:2007qw,Cai:2010zma,Avelino:2012ue,Barrow:2004ad,Haro:2015zda,Elizalde:2014uba,Das:2017jrl,
Cai:2008qw,Finelli:2001sr,Cai:2011ci,Haro:2015zta,Cai:2011zx,Haro:2014wha,Brandenberger:2009yt,deHaro:2014kxa,Odintsov:2014gea,
Qiu:2010ch,Bamba:2012ka,deHaro:2012xj,Nojiri:2019lqw,Elizalde:2020zcb,WilsonEwing:2012pu}. 
Beside producing an observationally compatible primordial power spectrum, 
the bouncing scenario has the merit to give rise to a singularity free evolution of the early universe. 
Although it has been argued that the Big-Bang singularity may be avoided through a suitable quantum generalization of gravity, 
the absence of a consistent quantum theory of gravity makes the bouncing description of the universe a promising scenario. 
In this context it may be metioned that string theory \cite{Horava:1995qa,Horava:1996ma,Polchinski:1998rq,Polchinski:1998rr} 
inherently incorporates the quantum nature of gravity and is associated with several extra spatial dimensions. Although originally 
intended to unify the known forces of nature, it turns out that extra dimensions can also provide plausible resolution to the 
gauge-hierarchy problem or the finetuning problem in particle physics arising due to large quantum corrections of the 
Higgs mass \cite{Antoniadis:1990ew,Horava:1995qa,Horava:1996ma,Lykken:1996fj,Kakushadze:1998wp,
ArkaniHamed:1998rs,Antoniadis:1998ig,ArkaniHamed:1998nn,Randall:1999ee,Randall:1999vf,Lykken:1999nb,Kaloper:1999sm}. 
In this context warped geometry models are particularly relevant. In particular, the warped geometry models due to 
Randall-Sundrum (RS) \cite{Randall:1999ee} earned a lot of attention since it resolves the gauge hierarchy problem without 
introducing any intermediate scale (between Planck and TeV scale) in the theory.

The RS scenario consists of an extra spatial dimension 
(over the usual four dimensional spacetime) with $S^1/Z_2$ orbifold symmetry, enclosed between two 3-branes which are considered to be flat. The distance $r_c$ between the two branes governs the magnitude of the brane warping and therefore plays the crucial role in resolving the gauge-hierarchy problem.
The assumption of flat branes which gives rise to a vanishing brane cosmological constant in the RS scenario can be relaxed in a generalized warped braneworld model \cite{Das:2007qn}, which allows the branes to be non-flat giving rise to de Sitter (dS) or anti-de Sitter (AdS) branes. The interplay of the brane warping (which depends on the interbrane distance $r_c$) and the magnitude of the brane cosmological constant leads to the resolution of the gauge-hierarchy problem in such models. The cosmological, astrophysical and phenomenological 
implications of warped braneworld models (with flat or non-flat branes) have been discussed in \cite{Csaki:2000zn,
DeWolfe:1999cp,Lesgourgues:2000tj,Csaki:1999mp,Binetruy:1999ut,Csaki:1999jh,Cline:1999yq,Nojiri:2000gv,Nojiri:2001ae,
Davoudiasl:1999jd,Das:2015zxa,Tang:2012pv,Arun:2014dga,Das:2013lqa,Banerjee:2018kcz,Chakraborty:2013ipa,
Paul:2016itm,Banerjee:2017lxi,Mitra:2017run,Elizalde:2018rmz,Aditya:2020thl,Banerjee:2019nnj,Banerjee:2019sae}. 
Since the resolution of the finetuning problem essentially depends on the interbrane distance $r_c$, the  
stabilization of $r_c$ to the appropriate value becomes crucial. The stabilization is achieved in the RS 
scenario by introducing a scalar field in the bulk \cite{Goldberger:1999uk,Goldberger:1999un}, which 
leads to a potential for $r_c$ in the four dimensional effective action whose minima can be suitably 
adjusted to address the finetuning problem. The origin of the bulk scalar however remains unexplored. 
This problem can evaded in the non-flat warped braneworld scenario with dynamical branes such that the 
interbrane distance attains the status of a field (the so called radion or the modulus). Such a framework 
enables the radion to generate its own potential along with a non-canonical kinetic term in the four 
dimensional effective action which in turn can stabilize the modulus to the suitable value \cite{Banerjee:2017jyk,Banerjee:2019nnj}, 
without invoking any additional scalar field in the theory.

The non-canonical scalar kinetic term becomes negative for certain values of the modulus which endows the 
radion a phantom-like behavior where the null energy condition is violated. Such a violation is a generic 
feature observed in a bouncing universe, which motivates us to explore the prospect of the non-flat warped 
braneworld model in addressing  bouncing cosmology. We investigate the cosmological evolution of the radion 
field in the FRW background and the subsequent evolution of the primordial fluctuations which allows us to 
understand the viability of the model in purview of the Planck 2018 constraints.

The paper is organized as follows: in \ref{S2}, we briefly describe the non-flat warped braneworld model and its four dimensional effective theory. 
Having set the stage, \ref{S3} is dedicated for studying the background cosmological evolution while the evolution of the perturbations and confrontation of the theoretical predictions with the latest Planck observations is discussed in \ref{S4}. We conclude with a summary of our results and a discussion of our findings in \ref{S5}.



\section{The non-flat warped braneworld scenario}
\label{S2}

Randall \& Sundrum (RS)\cite{Randall:1999ee} proposed the warped braneworld scenario to address the fine-tuning problem in particle physics. 
The RS model consists of a 5-dimensional AdS bulk bounded by two 3-branes, namely the visible brane (where our 4-d universe resides) and the 
hidden brane. The extra dimension denoted by $\phi$ is associated with a $S^1/Z_2$ orbifold symmetry and the hidden brane resides at $\phi=0$ 
while the visible brane is located at $\phi=\pi$. In the RS scenario, the bulk metric is described by  
\begin{align}
{ds}^2=e^{-2A(r_c, \phi)}\eta_{\mu \nu} {dx}^{\mu}{dx}^{\nu} -r_c ^2 {d\phi}^2 
\label{Eq1}
\end{align}
from which it is evident that the branes are flat which is ensured by the exact cancellation of the brane tension and the cosmological 
constant induced on the brane \cite{Sasaki:1999mi}.
In \ref{Eq1} the warp factor is represented by $e^{-2A}$ with $A=k_0r_c|\phi|$ where $r_c$ is the compactification radius and $k_0=\sqrt{-\Lambda/24M^3}$ 
such that $\Lambda$ and $M$ denote the five dimensional cosmological constant and Planck mass respectively.
The presence of the exponential warp factor in the metric ensures that $r_c \sim 12$ is sufficient to bring down the Higgs mass from the 
Planck scale to the TeV scale on the visible brane without introducing any new energy scale in the theory.
\\

A generalization of the RS model to incorporate non-flat branes is important since the non-flatness of our universe is often evident from the physical situations, e.g., an expanding universe, the presence of black holes etc. This is achieved by replacing the brane metric $\eta _{\mu \nu}$ with $g_{\mu \nu}$ in the above metric ansatz. This induces a non-zero cosmological constant $\Omega$ on the brane inherited from the bulk which can be both positive or negative. In the situation where the branes are de-Sitter, the warp factor is given by \cite{Das:2007qn}:
\begin{align}
e^{-A}=\omega \sinh\left(\ln\frac{c_2}{\omega}-k_0r_c |\phi| \right)
\label{Eq2}
\end{align}  
where, $\omega=(\Omega/3k_0^2)$ is a dimensionless constant directly proportional to the brane cosmological constant 
$\Omega$ and $c_2=1+\sqrt{1+\omega ^2}$.
It can be shown that the above warp factor can give rise to the requisite warping of the Higgs mass on the visible brane 
(i.e., $k_0 r_c \pi\sim 16\ln 10$) as in the RS scenario while keeping $\Omega \sim 10^{-124}$ (the magnitude of the present day 
cosmological constant in Planckian units). 

For AdS branes, it can be shown that the warp factor is given by $e^{-A'}=\omega cosh(ln\frac{\omega}{c_1}+k_0r_c |\phi|)$, 
with $c_1=1+\sqrt{1-\omega ^2}$ \cite{Das:2007qn}. We further note that in the event $\omega \rightarrow 0$, we retrieve 
the RS warp factor describing the flat braneworld scenario, for both the dS and AdS branes. 
Since the observed accelerated expansion of the universe can be explained by a positive brane cosmological constant, we 
will concentrate mainly on the warped braneworld scenario with de-Sitter branes and investigate its role in bouncing cosmology.

\subsection{The non-flat warped braneworld with the radion field}
\label{S2-1}
In the warped braneworld scenario, the resolution of the gauge hierarchy problem depends crucially on $r_c$, 
the stable distance between the two branes.
This requires a mechanism to stabilize the inter-brane distance to the suitable value. Goldberger \& Wise \cite{Goldberger:1999uk} 
addressed this by invoking a bulk scalar in the five dimensional action which resulted in a potential for $r_c$ in the 4-dimensional 
effective action. They showed that the stable value of $r_c$ corresponds to the minima of the potential. However, the physical origin of the scalar 
field in the bulk action is not well understood. 

Instead of flat branes if one considers the non-flat warped braneworld scenario, and allows the inter-brane distance to be treated as 
a 4-dimensional field $T(x)$ (the so called radion or the modulus), then it can be shown that a potential for the modulus is naturally 
generated in the 4-d effective action which in turn can stabilize the modulus \cite{Banerjee:2017jyk}. This modulus potential is 
completely attributed to the non-flat character of the branes and in the event the branes are flat this potential identically vanishes.  

This scenario is described by the bulk action,
\begin{align}
\mathcal{S}=\mathcal{S}_{gravity}+\mathcal{S}_{vis}+\mathcal{S}_{hid} \label{Eq3}
\end{align}
such that,
\begin{align}
\mathcal{S}_{gravity}=\int_{-\infty} ^{\infty} d^4x \int_{-\pi} ^{\pi} d\phi \sqrt{-G}(2M^3 \mathcal{ R} - \Lambda ) \label{Eq4}
\end{align}

\begin{align}
\mathcal{S}_{vis}=\int_{-\infty}^{\infty} d^4x \sqrt{-g_{vis}} (\mathcal{L}_{vis} -\mathcal{V}_{vis}) \label{Eq5}
\end{align}

\begin{align}
\mathcal{S}_{hid}=\int_{-\infty}^{\infty} d^4x  \sqrt{-g_{hid}}(\mathcal{L}_{hid} -\mathcal{V}_{hid})\label{Eq6}
\end{align}
where $\mathcal{R}$ is the bulk Ricci scalar and $G$ the determinant of the bulk metric $G_{\mu\nu}$. In \ref{Eq5}, 
$\mathcal{V}_{vis}$ and $\mathcal{L}_{vis}$ refer to the brane tension and matter Lagrangian on the visible 
brane $\phi=\pi$ while $\mathcal{V}_{hid}$ and $\mathcal{L}_{hid}$ in \ref{Eq6} correspond to the brane 
tension and matter Lagrangian on the hidden brane $\phi=0$.

The bulk is governed by the Einstein's equations with the following solution for the metric,
\begin{align}
{ds}^2=e^{-2A(x,\phi)}g_{\mu \nu} {dx}^{\mu}{dx}^{\nu} - T(x)^2 {d\phi}^2.  \label{Eq7}
\end{align}
which is easily extended from \ref{Eq1} with $\eta_{\mu\nu}$ replaced by $g_{\mu\nu}$ and $r_c$ substituted by the radion field $T(x)$. 
We will concentrate on de-Sitter branes in this work and hence the form of the warp factor is given by
\begin{align}
e^{-A}=\omega \sinh\left(\ln\frac{c_2}{\omega}-k_0T(x) |\phi| \right)
\label{Eq8}
\end{align} 
which is the same as \ref{Eq2} with $r_c$ replaced by $T(x)$. 

It is interesting to note that the the positivity of the warp factor in \ref{Eq8} requires that $\xi=(\Phi/f)=\exp\{-k_0T(x)\pi\}\geq (\omega/c_2)$ 
which immediately follows when we write the warp factor in the following way, 
\begin{align}
e^{-A}=\frac{\omega}{2}\left\{\exp \left[\left(\ln\frac{c_2}{\omega}-k_0T(x) |\phi| \right) \right]-\exp \left[-\left(\ln\frac{c_2}{\omega}-k_0T(x) |\phi| \right)\right] \right\}
=\frac{c_{2}}{2}\exp(-k_0 T(x)|\phi|)
-\frac{\omega ^{2}}{2c_{2}}\exp(k_0 T(x)|\phi|) \label{Eq9-1}
\end{align}
and demand $e^{-A}\geq 0$. Further, since the modulus $T(x)$ cannot be negative the maximum value 
that $\xi$ can attain is unity, when $T(x)\rightarrow 0$. Therefore, throughout this work our region of 
interest in the field space would be $\omega/c_2 \leq \xi \leq 1$. We will use this property of 
the warp factor when we explore bouncing cosmolgy with the radion field in \ref{S3}.

The effective action $S$ in four dimensions is derived from the bulk action $\mathcal{S}$ by integrating over the extra coordinate $\phi$. 
This can be segregated into three parts, namely, 
\begin{align}
~S=~S_{1}+ ~S_2 + ~S_3 
\label{Eq9}
\end{align}
where,
\begin{align}
S_1=\frac{2M^{3}}{k_0}\int d^4 x~\sqrt{-\hat{g}}~h\left(\frac{\Phi}{f}\right)\hat{R}~
\label{Eq10}
\end{align} 
is the curvature dependent part of the effective action $S$ with $\hat{g}$ the determinant and $\hat{R}$ the Ricci 
scalar with respect to the brane metric $\hat{g}_{\mu\nu}$. From \ref{Eq10}  it is evident that the Ricci scalar 
involves a coupling with the dimensionless radion field $\xi=\Phi/f\equiv \exp\{-k_0 T(x)\pi\}$ with $f=\sqrt{6M^3c_2^2/k_0}$ 
and hence is in the Jordan frame. Here and in the rest of the discussion we shall denote $\xi$ as the radion field. 
The coupling of the modulus to $\hat{R}$ is denoted by $h(\xi)$ which assumes the form,
\begin{align}
h\left(\xi\right)=\left\{\frac{c_2^2}{4}+\omega^2 \ln \xi+\frac{\omega^4}{4c_2^2}\left(\frac{1}{\xi^2}\right)-\frac{\omega^4}{4c_2^2}-\frac{c_2^2}{4}\xi ^2 \right\}
\label{Eq11}
\end{align}
The second part of the effective action $S_2$ comprises of a potential for the radion field,
\begin{align}
~S_{2}=-2M^3 k_0\int d^4x~\sqrt{-\hat{g}}~\hat{V}\left(\xi\right) 
\label{Eq12}
\end{align}
with 
\begin{align}
\hat{V}\left(\xi\right)=6\omega^4 \ln \xi-\frac{3}{2}\omega^2c_2^2 \xi^2 +\frac{3}{2}\omega^2c_2^2+\frac{3}{2}\frac{\omega^6}{c_2^2}\left(\frac{1}{\xi^2}\right)-\frac{3}{2}\frac{\omega^6}{c_2^2} =6\omega ^2  h\left(\xi\right)
\label{Eq13}
\end{align}
It is interesting to note that the potential $\hat{V}(\xi)$ is directly proportional to $h(\xi)$ and vanishes in the event the 
branes are flat i.e., $\omega\rightarrow 0$ \cite{Goldberger:1999un}.
Moreover, it has an inflection point at $\xi_i=\omega/c_2$ which will have important consequences when we explore early universe cosmology in this model.  

The third part of the effective action in \ref{Eq9} is associated with the kinetic term of the radion given by,
\begin{align}
~S_3=\int d^4x~\sqrt{-g}~\left(\frac{1}{2}\partial_\mu \Phi \partial^\mu \Phi\right)
\hat{G}\left(\xi\right)
\label{Eq14}
\end{align}
where,
\begin{align}
\hat{G}\left(\xi\right)=1+\frac{4}{3}\frac{\omega^2}{c_2^2}\left(\frac{1}{\xi^2}\right)\ln \xi -\frac{\omega^4}{c_2^4}\left(\frac{1}{\xi^4}\right) 
\label{Eq15}
\end{align}
denotes the non-canonical coupling to the kinetic term which reduces to the canonical form when $\omega\rightarrow 0$. 
Therefore, the non-flatness of the branes generates the brane cosmological constant $\Omega$ which in turn gives rise 
to the potential for the radion and its non-canonical kinetic term.

\begin{figure*}[t!]
\centering
\subfloat[\label{Fig_2a}]{\includegraphics[scale=0.64]{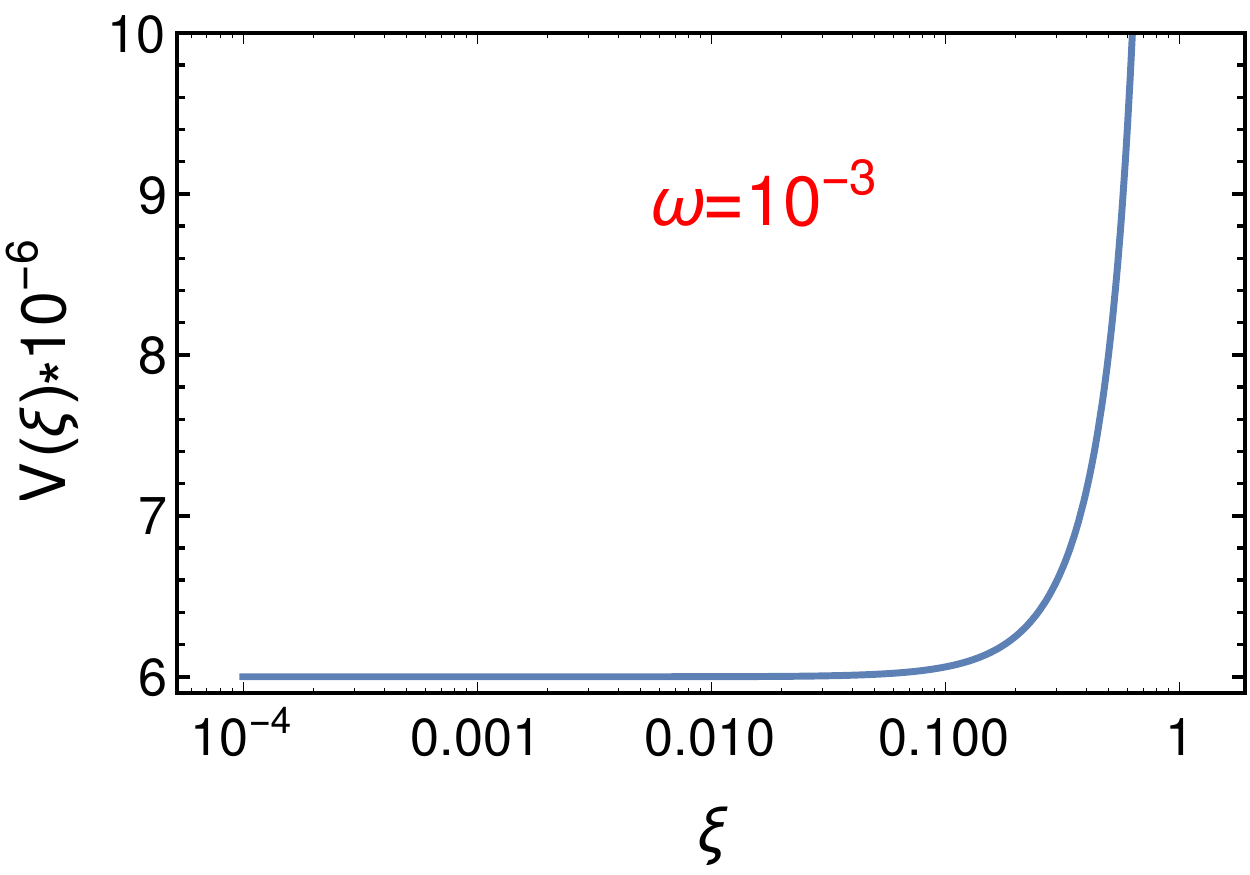}}~~
\subfloat[\label{Fig_2b}]{\includegraphics[scale=0.64]{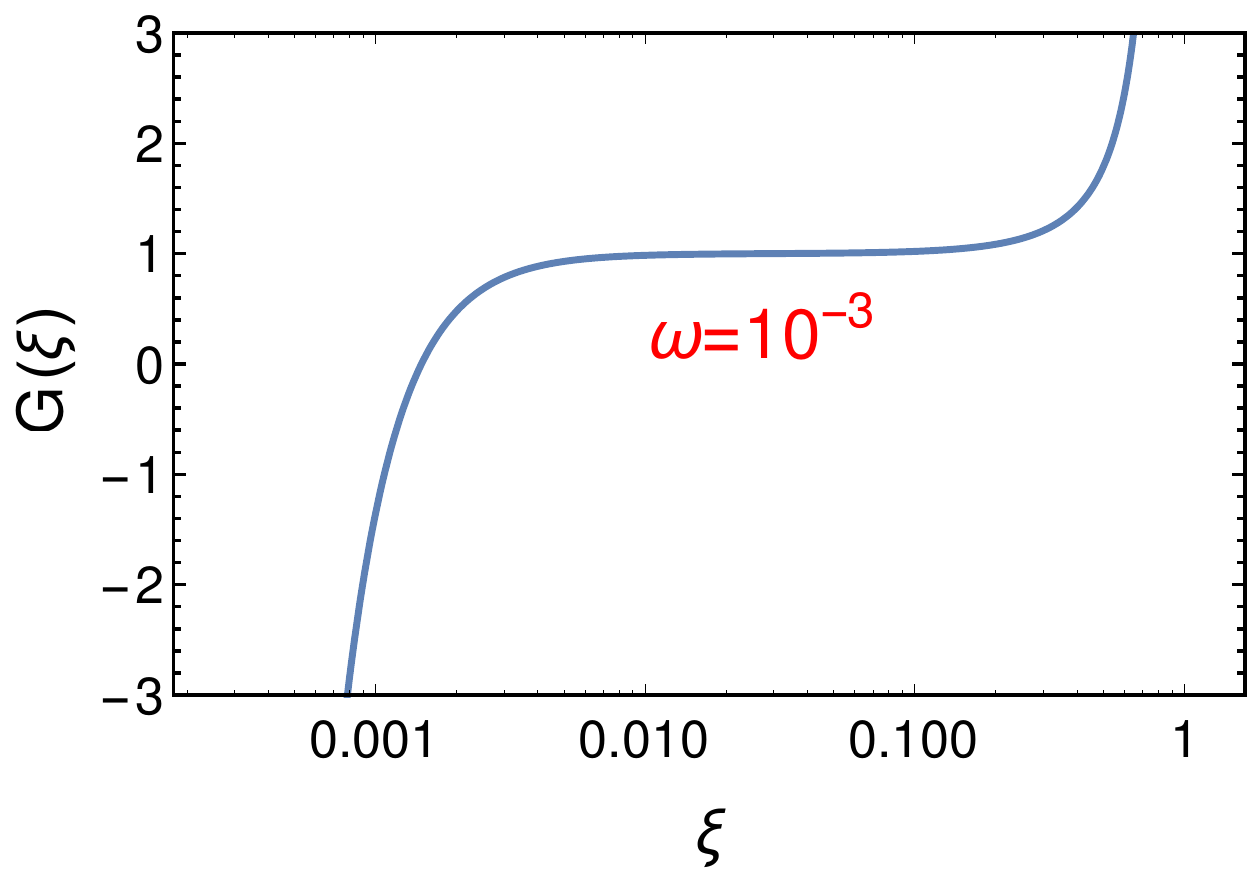}}
\caption{The above figure depicts the variation of (a) the radion potential $V$ and (b) the non-canonical 
coupling to the kinetic term $G$ in the Eintein frame, within the allowed range of the radion field $\xi$ for $\omega=10^{-3}$.}
\label{Fig_02}
\end{figure*}

Since the observations are generally made in the Einstein frame, we perform a conformal transformation of the 
Jordan frame metric $\hat{g}_{\mu\nu}$ to remove the coupling of the scalar field to the Ricci scalar. 
This is achieved by scaling the Jordan frame metric $\hat{g}_{\mu \nu}$ with the conformal field $\zeta(x)$ 
such that the metric in the Einstein frame is given by $g_{\mu\nu}=\zeta^2 (x)\hat{g}_{\mu \nu}$. 
With this conformal scaling it can be shown that in four dimensions, 
the  Ricci scalar $R$ in the Einstein frame is related to the Ricci scalar $\hat{R}$ in the Jordan frame by,
\begin{align}
R=\left[\frac{\hat{R}}{\zeta^2}
-\frac{6}{\zeta^3}\hat{g}^{\mu\nu}\hat{\nabla}_\nu\hat{\nabla}_\mu\zeta \right]
\label{Eq16}
\end{align}
where $\hat{\nabla}$ denotes covariant derivative with respect to the metric $\hat{g}_{\mu\nu}$.
Using \ref{Eq16} and the fact that $\sqrt{-\hat{g}}=\zeta^{-4}\sqrt{-g}$ and choosing  $\zeta\equiv \sqrt{h(\Phi/f)}$, 
we arrive at the effective action in the Einstein frame, 
\begin{align}
\mathcal{A}=\int d^4 x \sqrt{-g}\Bigg[\frac{R}{2 \kappa^2}
+\frac{1}{2}G(\xi)\partial^\mu\Phi\partial_\mu\Phi 
-8 M^3\kappa^2 V(\xi)\Bigg]  
\label{Eq17} 
\end{align}
where $2\kappa^2=16 \pi G_N=\frac{k_0}{2M^3}$ and the potential due to the radion field in the Einstein grame is given by,
\begin{align}
V(\xi)=\frac{\hat{V}(\xi)}{h(\xi)^2}=\frac{6\omega^2}{h(\xi)}
\label{Eq18} 
\end{align}
while the non-canonical coupling to the kinetic term is given by,
\begin{align}
G(\xi )&=\frac{\hat{G}(\xi)}{h(\xi)}+\frac{1}{c_2^2}\bigg[\frac{h^\prime(\xi)}{h(\xi)}\bigg]^2  
\label{Eq19}   
\end{align}
where `prime' here implies differentiation with respect to $\xi$.
In \ref{Fig_2a} and \ref{Fig_2b} we plot the variation of $V$ and $G$ with the radion field $\xi$ for $\omega=10^{-3}$. 
It can be shown from \ref{Eq18} that the radion potential $V$ in the Einstein frame continues to have an inflection 
point at $\xi_i=\omega/c_2$ which can be confirmed from the vanishing first and second derivatives but a positive 
third derivative of $V(\xi)$ with respect to $\xi$ at $\omega/c_2$. Moreover, \ref{Fig_2b} reveals that the non-canonical 
coupling to the kinetic term $G(\xi)$ exhibits a transition from a normal to a phantom regime (i.e from $G(\xi) > 0$ to $G(\xi) < 0$), 
where the phantom like behavior remains when $\xi$ lies in the range $\xi_i\leq \xi \leq\xi_f$, 
with $\xi_f$ denoting the zero crossing of $G(\xi)$. Also note from \ref{Fig_2b} that for $\omega = 10^{-3}$, $\xi_f \simeq 0.00148$.

We thus note that in the non-flat warped braneworld scenario, we have a scalar field, the radion, 
which is associated with a potential and a non-canonical kinetic term. It is believed that in the early universe 
the big bang singularity can be avoided in a bouncing scenario which is triggered by a scalar field with a potential. 
This raises the question whether the radion field can be instrumental in giving rise to a bouncing universe which we address in the next section.

\section{Implications in Early Universe Cosmology: Background evolution}
\label{S3}
In this section we explore the role of the radion field in triggering a bouncing universe which can potentially avoid the big bang singularity. 
The bouncing scenario often invokes a scalar field with a potential and there exist plenty of models in the literature which can 
give rise to such a scenario (see \cite{Battefeld:2014uga}). In most of the cases the scalar potentials are reconstructed to explain the observations 
and often their origin remains unexplained. The merit of the non-flat warped braneworld model lies in the fact that the the radion 
field naturally arises from compactification in the effective four-dimensional theory and generates its own potential and non-canonical 
kinetic term. Here we consider the implications of the radion field in inducing a bouncing universe.

Since we are interested to study early universe cosmology with the radion field we consider the metric in the Einstein frame to be 
described by the FRW spacetime in the spatially flat form,
\begin{align}
ds^2=dt^2-a(t)^2\bigg[dx^2+dy^2+dz^2\bigg] \label{Eq23} 
\end{align}
with $a(t)$ is known as the scale factor of the universe. In the generalized RS scenario, the 3-branes can be Minkowskian, de-Sitter 
or anti de-Sitter depending on the values of the induced brane cosmological constant \cite{Das:2007qn}. 
Recall, in the present work, we consider the branes to be de-Sitter, 
i.e our visible universe is a 3-brane (i.e having three spatial dimension along with the time coordinate), described by the spatially flat FRW metric. 
Since the FRW metric is curved irrespective of its spatial curvature, the visible 3-brane is non-flat. 
At this stage it deserves mentioning that the spatially flat FRW metric is more consistent over the 
closed or open FRW universe from latest Planck 2018 data through TT, TE, EE + lowE + lensing + BAO data, 
where TT means temperature temperature cross-correlation of CMB data,
TE means cross-correlation between temperature and electric type polarization of CMB data and finally BAO stands for 
Baryon Acoustic Oscillation \cite{Aghanim:2018eyx}. Due to the time dependency of the scale factor, 
the metric in \ref{Eq23} indicates a non-zero curvature on the four dimensional brane geometry and moreover the brane curvature is characterized 
by the corresponding Ricci scalar given by $R = 6\frac{\ddot{a}}{a} + 6\frac{\dot{a}^2}{a^2}$. As we will see from \ref{Eq27} that the kinetic 
as well as the potential energy of radion field contributes to the on-brane Ricci scalar via the effective four dimensional Freidmann equation. 
Furthermore, as shown in \ref{Eq18}, the potential energy of the radion field is proportional to the induced brane cosmological constant 
and thus one may argue that 
the radion potential energy is generated entirely due to the presence of the non-zero brane cosmological constant. Thereby the brane cosmological constant 
affects the evolution of the scale factor through the potential energy density of the radion field. Below, we will show that the 
metric ansatz of \ref{Eq23} is consistent with the field equations of motion and moreover it will lead to a non-singular bounce on our visible brane.

It is evident from \ref{Eq17} that the energy momentum tensor $T^{\mu}_{\nu}$ due to the radion field is given by,
\begin{align}
T^\mu_\nu=G(\xi)\partial^\mu \Phi \partial_\nu \Phi 
-\frac{1}{2}\delta^\mu_\nu G(\xi) \partial^\alpha \Phi \partial_\alpha \Phi + 2M^3k_0V(\xi) \delta^\mu_\nu
\label{Eq24}
\end{align}
such that
\begin{align}
T^0_0(\xi)=\frac{3M^3c_2^2}{k_0}G(\xi)\dot{\xi}^2 + 2M^3 k_0 V(\xi)=\rho
\label{Eq25}
\end{align}
represents the energy density while  
\begin{align}
-T^i_j(\xi)=\delta^i_j \bigg[\frac{3M^3c_2^2}{k_0} G(\xi)\dot{\xi}^2 - 2M^3k_0 V(\xi)\bigg]=p
\label{Eq26}
\end{align}
corresponds to the pressure due to the radion field. We note that the radion field $\xi$ depends only on time since 
the background metric given by \ref{Eq23} is only time dependent.

Using \ref{Eq25} the Friedmann equation obtained from the temporal component of the Einstein's equations assume the form,
\begin{align}
H^2 =\frac{\kappa^2}{3}\rho(t) =\frac{c_2^2}{4} G(\xi) \dot{\xi}^2 + \frac{k_0^2}{6} V(\xi)
\label{Eq27}
\end{align}
while the Friedman equation derived from the spatial component of the Einstein's equations is given by,
\begin{align}
\dot{H} =-\frac{\kappa^2}{2}(\rho + p) = -\frac{3}{4}c_2^2 G(\xi) \dot{\xi}^2
\label{Eq28}
\end{align}
where $H=\dot{a}/a$ denotes the Hubble parameter.
The equation of motion for the radion field is given by,
\begin{align}
\label{Eq29}
\dot{\rho}+3H(\rho+p)=0
\end{align}
Using \ref{Eq25} and \ref{Eq26}, \ref{Eq29} can be written as,
\begin{align}
\ddot{\xi} + 3H \dot{\xi} + \frac{G^\prime(\xi)}{2G(\xi)}\dot{\xi}^2 + \frac{k_0^2}{3c_2^2}\frac{V^\prime(\xi)}{G(\xi)}=0
\label{Eq30}
\end{align}
\ref{Eq27}, \ref{Eq28} and \ref{Eq30} are the background equations, although it is important to note that \ref{Eq30} 
is not independent but can be derived from \ref{Eq27} and \ref{Eq28}.\\ 
At this stage it deserves mentioning that in a canonical scalar tensor theory, the 
Friedmann equation becomes $\dot{H} \propto -\dot{\xi}^2$ (\ref{Eq28}) and thus the Hubble parameter decreases monotonically with cosmic time. Therefore a bounce phenomena is impossible in a canonical scalar tensor model as it cannot give rise to $\dot{H} > 0$ which is one of the necessary conditions 
to get a bounce. On the contrary, in a non-canonical scalar tensor theory where the scalar field has non-canonical kinetic term (as $G(\xi)$ in 
the present context), the Friedmann equations are modified due to the presence of $G(\xi)$ and the modified equations are given by 
\ref{Eq27} and \ref{Eq28} respectively. \ref{Eq28} clearly indicates that in a non-canonical scalar tensor model, the sign of $G(\xi)$ actually 
controls the energy condition, in particular $G(\xi) < 0$ leads to a violation of null energy condition which in turn may ensure a bouncing 
phase in our visible universe. We have already noted in \ref{S3} that in the present scalar-tensor model the non-canonical kinetic term exhibits a transition 
from a normal to a phantom regime (i.e from $G(\xi) > 0$ to $G(\xi) < 0$) where the null energy condition is violated. Therefore it is important to investigate the prospect of bouncing cosmology with the present non-flat warped braneworld model which we explore next. In particular, we first present the background evolution of $H(t)$ and $\xi(t)$ (governed by \ref{Eq27} and \ref{Eq28}) 
in the next section and subsequently study the evolution of the perturbations in \ref{S4}.


In general, a non-singular bounce is characterized by the conditions $H(t_b) = 0$ and $\dot{H}(t_b) > 0$ where $t_b$ is the cosmic time when 
the bounce occurs. Keeping these conditions in mind, if we look into \ref{Eq27} and \ref{Eq28}, then it is evident that the model has a possibility to show 
a bounce phenomena when the non-canonical function $G(\xi)$ becomes negative i.e when the radion field is in the phantom regime. The analysis in
\ref{S3} reveals that $G(\xi)$ is indeed negative in the regime $\xi \sim \omega$. Thus, at first we analytically 
solve the background equations near $\xi \sim \omega$ to investigate the bounce and then we numerically determine 
the background evolution for a wide range of $\xi$ (or equivalently for a wide range 
of cosmic time), where the boundary conditions of the numerical calculation are provided from the previously found analytic solutions.\\
In particular we consider,
\begin{align}
\xi(t)=\frac{\omega}{c_2} [1+ \delta(t)]
\label{Eq31}
\end{align}
with $\delta(t) \ll 1$. Due to the above form of $\xi(t)$, $h(\xi)$ in \ref{Eq11} simplifies to
\begin{align}
h(\xi)=\frac{c_2^2}{4} + \mathcal{O}(\omega^2) \simeq \frac{c_2^2}{4}
\label{Eq32}
\end{align}
such that $V(\xi)$ is given by,
\begin{align}
V(\xi) \simeq \frac{24\omega^2}{c_2^2}
\label{Eq33}
\end{align}
while $G(\xi)$ can be approximated as,
\begin{align}
G(\xi)\simeq -\frac{16}{3c_2^2}\bigg[ ln \bigg(\frac{c_2}{\omega}\bigg)-\bigg \lbrace 4 + 2 ln \bigg(\frac{c_2}{\omega}\bigg
)\bigg\rbrace \delta \bigg]
\label{Eq34}
\end{align}
The above simplifications in the form of $V(\xi)$ and $G(\xi)$ hold only 
in the regime where $\xi(t)$ is given by \ref{Eq31} with $\delta(t)\ll 1$. 
With these simplifications the evolution equations for the Hubble parameter $H(t)$ and the radion field (i.e \ref{Eq27} and \ref{Eq28}) turn out to be,
\begin{align}
\dot{H} + 3H^2 -\frac{12 k_0^2 \omega^2}{c_2^2}=0
\label{Eq35}
\end{align} 
and
\begin{align}
\dot{\delta}^2=\frac{c_2^2}{\omega^2}\frac{\dot{H}}{4 ln \big(\frac{c_2}{\omega}\big)}\bigg[1+  \delta \bigg\lbrace \frac{4+2ln\big(\frac{c_2}{\omega}\big)}{ln\big(\frac{c_2}{\omega}\big)}\bigg\rbrace \bigg]
\label{Eq36}
\end{align}
respectively. \ref{Eq35} can be solved to obtain the time evolution of the Hubble parameter which assumes the form,
\begin{align}
H(t)=2k_0\frac{\omega}{c_2}tanh \bigg[6 \frac{\omega}{c_2}k_0 t\bigg]
\label{Eq37}
\end{align}
such that $\dot{H}$ is given by,
\begin{align}
\dot{H}=12 k_0^2\frac{\omega^2}{c_2^2} sech^2\bigg[6 \frac{\omega}{c_2}k_0 t\bigg]
\label{Eq38}
\end{align}
Using \ref{Eq38} in \ref{Eq36} we obtain,
\begin{align}
\dot{\delta}=- k_0 \sqrt{\frac{3}{ln\big(\frac{c_2}{\omega}\big)}} \bigg(1 + \frac{A\delta}{2} \bigg) sech\bigg[\frac{6\omega}{c_2}k_0 t   \bigg]
\label{Eq39}
\end{align}
with $A=\frac{4+2ln \big(\frac{c_2}{\omega}\big)}{ln\big(\frac{c_2}{\omega}\big)}$ and solving the above equation yields 
the following time evolution for $\delta(t)$,
\begin{align}
\delta(t)=-\frac{2}{A} + C_1 exp \bigg[ -\frac{A}{6} \frac{\omega}{c_2}\sqrt{\frac{3}{ln\big(\frac{c_2}{\omega}\big)}}tan^{-1} tanh \bigg( \frac{3\omega}{c_2}k_0 t \bigg)  \bigg]
\label{Eq41}
\end{align}
where $C_1$ is an integration constant which can be determined by demanding,
\begin{align}
\lim_{t \to \infty} \delta(t) \to 0~~~~~~~~~\mathrm{or~equivalently}~~~~~~~~~~\lim_{t\to \infty} \xi(t) \to \frac{\omega}{c_2}
\label{Eq42}
\end{align}
which implies that $\xi(t)$ (i.e the radion field) monotonically decreases with time and asymptotically goes to the value 
$\frac{\omega}{c_2}$ which is the minimum possible value of $\xi$ from the requirement of positive warp factor, as discussed after \ref{Eq9-1}. 
From the condition given by \ref{Eq42}, the form of $C_1$ turns out to be,
\begin{align}
C_1=\frac{2}{A}exp\bigg[\frac{A}{6}\frac{\omega}{c_2} \sqrt{\frac{3}{ln \big( \frac{c_2}{\omega}\big)}} \frac{\pi}{4} \bigg]
\label{Eq44}
\end{align}
which when substituted in \ref{Eq41} gives,
\begin{align}
\delta(t)=\frac{2}{A}\bigg[exp \bigg\lbrace -\frac{A}{6}\frac{\omega}{c_2} \sqrt{\frac{3}{ln\big( \frac{c_2}{\omega} \big)}}\bigg ( tan^{-1} tanh \bigg(\frac{3\omega}{c_2}k_0 t\bigg) -\frac{\pi}{4}\bigg)   \bigg \rbrace -1  \bigg]
\label{Eq45}
\end{align}
With this, we arrive at the background solution for $H(t)$ and $\delta(t)$ in the regime $\xi(t)=\frac{\omega}{c_2}(1+\delta(t))$ with $\delta(t)\ll 1$,
\begin{align}
\xi(t)=\frac{\omega}{c_2}(1+\delta(t)),
\begin{cases}
H(t)=2k_0\frac{\omega}{c_2}tanh \bigg[6 \frac{\omega}{c_2}k_0 t\bigg] & \\
    \delta(t)=\frac{2}{A}\bigg[exp \bigg\lbrace -\frac{A}{6}\frac{\omega}{c_2} \sqrt{\frac{3}{ln\big( \frac{c_2}{\omega} \big)}}\bigg ( tan^{-1} tanh \bigg(\frac{3\omega}{c_2}k_0 t\bigg) -\frac{\pi}{4}\bigg)   \bigg \rbrace -1  \bigg],              
\end{cases}
\label{Eq46}
\end{align}
where, $A=\frac{4+2ln \big(\frac{c_2}{\omega}\big)}{ln\big(\frac{c_2}{\omega}\big)}$. \ref{Eq46} 
clearly indicates $H(0) = 0$ and $\dot{H} > 0$ at $t=0$ (corresponding to the bounce time) which 
are the necessary conditions for a non-singular bounce. Therefore, the present non-flat warped braneworld 
model predicts a bouncing universe in the visible brane 
when the radion field lies within the phantom regime, in particular near $\xi \sim \omega$. In the phantom regime, due to the negative 
kinetic energy of the radion field, the effective null energy condition (NEC) is violated and makes the bounce possible 
at a certain finite time, in particular at $t = 0$. Here we would like to mention that such NEC violation occurs irrespective of any 
values of $\frac{\omega}{c_2}$ and $k_0$ (i.e the model parameters), and moreover the initial conditions of the solution of 
\ref{Eq46} is free from fine tuning of the model parameters. Thereby, we may argue that the bounce solution in the present context 
is a generic feature and does not require any fine-tuned values of the model parameters.

\begin{figure*}[t!]
\centering
\subfloat[\label{Fig_3a}]{\includegraphics[scale=0.69]{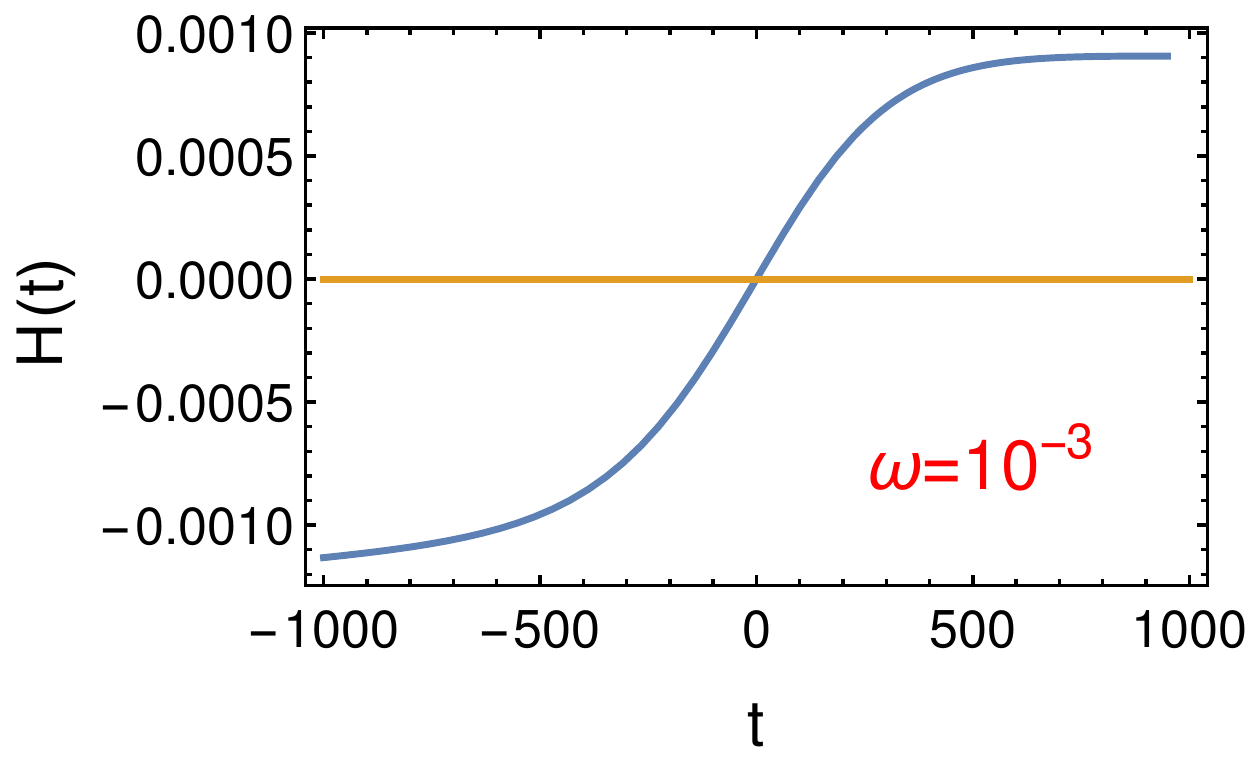}}~~
\subfloat[\label{Fig_3b}]{\includegraphics[scale=0.64]{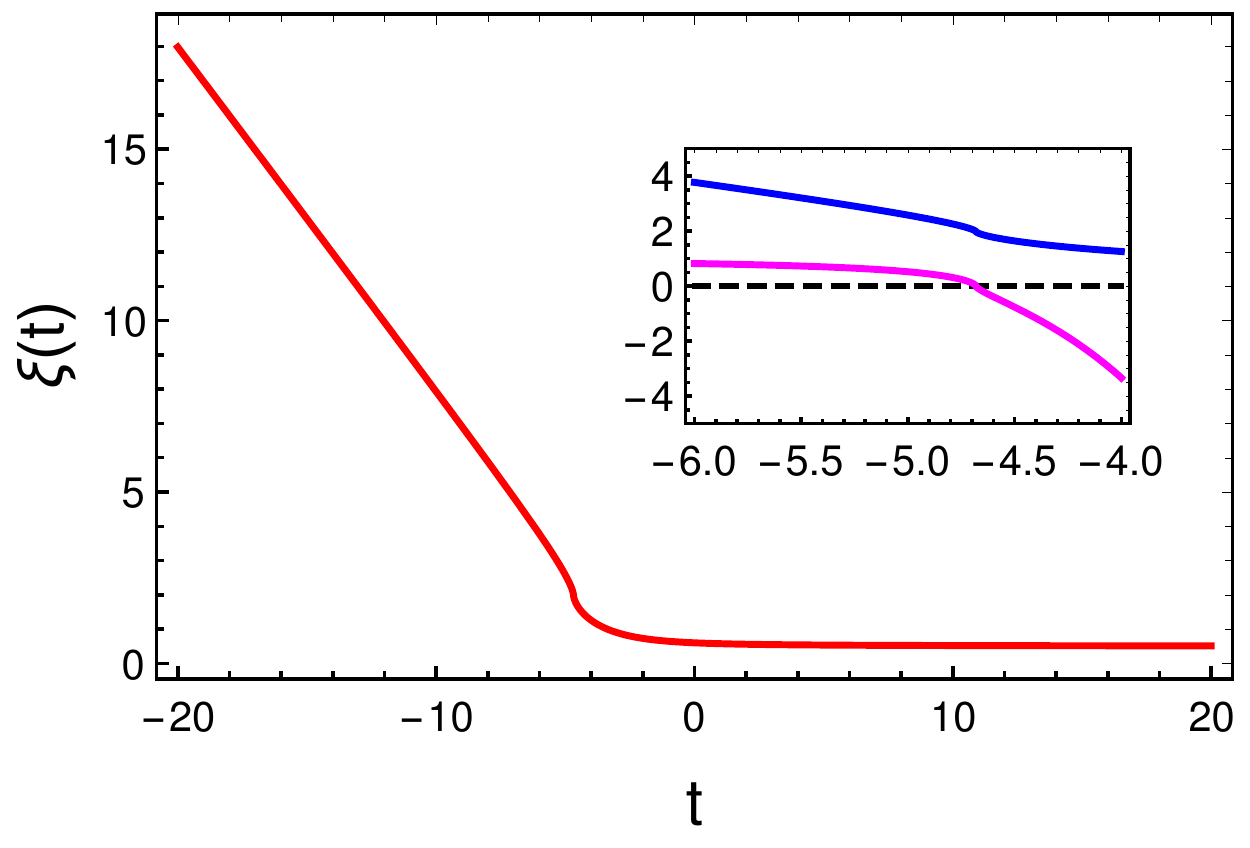}}
\caption{The above figure depicts the time evolution of (a) the Hubble parameter $H(t)$ and 
(b) the radion field magnified 1000 times, i.e. $\xi(t)\times1000$; while the inset of \ref{Fig_3b} 
depicts the non-canonical kinetic term $G(\xi)$ (magenta curve) 
and the zoomed-in version of $\xi(t)\times1000$ (blue curve) near the zero crossing of $G(\xi)$. 
Note that bounce occurs at $t=0$ when the kinetic term of the radion is in the phantom regime. 
Both the above figures are illustrated for $\omega=10^{-3}$.}
\label{Fig_03}
\end{figure*}

At this stage, it is important to check 
whether the radion field, starting from a value in the normal regime, will reach to the phantom regime by its $dynamical~evolution$. For this purpose, 
we solve the coupled equations for $H(t)$ and $\xi(t)$ (i.e \ref{Eq27} and \ref{Eq28}) for a wide range of cosmic time numerically. 
In regard to the numerical calculation, the boundary conditions are provided 
from the analytic solutions as determined in \ref{Eq46}, in particular the boundary conditions are given by $H(0) = 0$ and 
$\xi(0) = 6.0041\times10^{-4}$, where we consider 
$\omega = 10^{-3}$ (later, during the perturbation calculation, we show that such a value of $\omega$ is consistent with 
the Planck 2018 constraints). The time evolution of the Hubble parameter and the radion field are shown in \ref{Fig_3a} and \ref{Fig_3b} respectively 
(the radion field plot is magnified 1000 times i.e $\xi(t)\times1000$). 
In the inset of \ref{Fig_3b}, 
the magenta curve denotes the time evolution of $G(\xi)$ while the blue curve represents the zoomed-in version of $\xi(t)\times1000$ 
near the zero crossing of $G(\xi)$. 
From \ref{Fig_3b} it is evident that $G(\xi)$ exhibits a transition from a normal regime 
(where $G(\xi) > 0$) to a phantom regime (where $G(\xi) < 0$) 
with its zero crossing occurs at a finite time before the bounce at $t = 0$. Moreover \ref{Fig_3b} demonstrates that 
there is no divergence in the dynamical evolution of $\xi(t)$ as $G(\xi)$ transits from normal to the phantom regime. 
However on the other hand, as evident from \ref{Eq28}, $\frac{d\xi}{dt} = \dot{\xi}$ 
diverges at the time when the non-canonical kinetic coupling $G(\xi)$ makes the zero crossing. Here we would like to mention that such divergence of 
$\dot{\xi}$ does not lead to any pathology to the radion field equation of motion i.e to \ref{Eq30} and the reason is following: \ref{Eq30} can be 
equivalently expressed as $\frac{d}{dt}\bigg(\frac{3M^3c_2^2}{k_0}G(\xi)\dot{\xi}^2 + 2M^3 k_0 V(\xi)\bigg) + \frac{18M^3c_2^2}{k_0}HG(\xi)\dot{\xi}^2 = 0$ which 
includes $G(\xi)\dot{\xi}^2$ and its derivative with respect to cosmic time. Now \ref{Eq28} evidents that $G(\xi)\dot{\xi}^2$ is proportional to 
$\dot{H}$ which, along with its derivative, is indeed finite for all possible cosmic time (see \ref{Fig_3a}). 
Thereby the term $G(\xi)\dot{\xi}^2$ and its derivative with respect to $t$ are finite everywhere even at $G(\xi) \rightarrow 0$, 
and thus we may argue that the radion field equation of motion does not lead to any inconsistency in the present context.\\

\ref{Fig_3a} reveals that the Hubble parameter becomes zero and increases with respect to cosmic time at $t = 0$, which confirms a non-singular bounce 
at $t = 0$. Before demonstrating the dynamics of the radion field, we recall that 
for $\omega = 10^{-3}$, the zero crossing of $G(\xi)$ occurs at $\xi=\xi_f \simeq 0.00148$, as shown in \ref{Fig_2b} i.e. 
$G(\xi)$ exhibits the normal to phantom transition as $\xi$ crosses $\xi_f = 0.00148$ from higher values. Numerical solution of \ref{Eq27} and \ref{Eq28} indicates that the radion field starts its journey from the normal regime (i.e $\xi > \xi_f$) and dynamically moves to the phantom era (i.e $\xi < \xi_f$) with time by monotonically decreasing in magnitude and asymptotically stabilizes to the value $\xi_i \to \omega/c_2$ which for $\omega = 10^{-3}  \sim \xi_i =5\times10^{-4}$. This is in accordance with the analytical results obtained in \ref{Eq46} and the time evolution of the background radion field is explicitly illustrated in \ref{Fig_3b}.
As the radion field asymptotically tends to $\xi_i$ (i.e., $\xi \to \xi_i+\varepsilon$), the warp factor $e^{-A}\simeq c_2\varepsilon$ which in turn resolves the gauge-hierarchy problem for $\varepsilon\simeq 10^{-16}$ while the stabilized inter-brane separation  
$k_0\pi\langle T \rangle \to \ln{\big(\frac{c_2}{\omega}\big)}$ \cite{Banerjee:2017jyk}. Therefore, in the non-flat warped braneworld scenario with dynamical branes, the radion generates its own potential which in turn stabilizes the modulus dynamically in the FRW background. Further, the presence of the phantom era enables violation of the null energy condition for the radion field which makes this a promising model to explore the bouncing scenaio.

At this stage, it may be mentioned that a holonomy improved non-canonical scalar tensor model may rescue the energy condition in a bouncing scenario. In the holonomy generalized model, the squared Hubble parameter (i.e $H^2$ in \ref{Eq27}) is proportional to the linear as well as quadratic power of 
energy density, unlike the usual Friedmann equations where $H^2$ is proportional only to the linear power of energy density. Such difference 
in the field equations may play a significant role to rescue the null energy condition necessary for a non-singular bounce. 
This investigation is expected to be carried out soon in a future work.

\section{Implications in Early Universe Cosmology: Evolution of perturbations}
\label{S4}
In this section, we consider the spacetime perturbations over the background FRW metric and consequently determine the primordial 
observable quantities like the scalar spectral index ($n_s$), tensor to scalar ratio ($r$) and the amplitude of scalar perturbations ($A_s$). 
In a bouncing universe, the Hubble parameter becomes zero and consequently the comoving Hubble radius diverges at the bouncing point. However, the asymptotic 
behaviour of the Hubble radius differentiates various bouncing models which can be broadly classified into two scenarios. In the first case, the comoving Hubble radius decreases and goes to zero asymptotically with time, 
which corresponds to a late time accelerating universe. In this case, the perturbation modes generate near the bounce, because at that time, the horizon has 
an infinite size and all the perturbation modes lie within the horizon. In the second situation the Hubble radius diverges asymptotically with time, which indicates a decelerating universe at late time and consequently 
the primordial perturbation modes relevant for the present era 
generate at a distant past far away from the bounce. More explicitly, in the latter case, the comoving wave number $k$ begins 
its journey from the infinite past in the contracting universe, within the sub-Hubble scale, exits the horizon as it contracts, and again re-enters the horizon in the low curvature regime of the expanding phase and becomes relevant for present time 
observations. 
Therefore, depending on the 
asymptotic behavior of the Hubble radius, the perturbation modes in a bounce model 
generate either near the bounce or far away from the bounce deeply in the contracting regime.

\begin{figure*}[t!]
\centering
\subfloat[\label{Fig_newa}]{\includegraphics[scale=0.66]{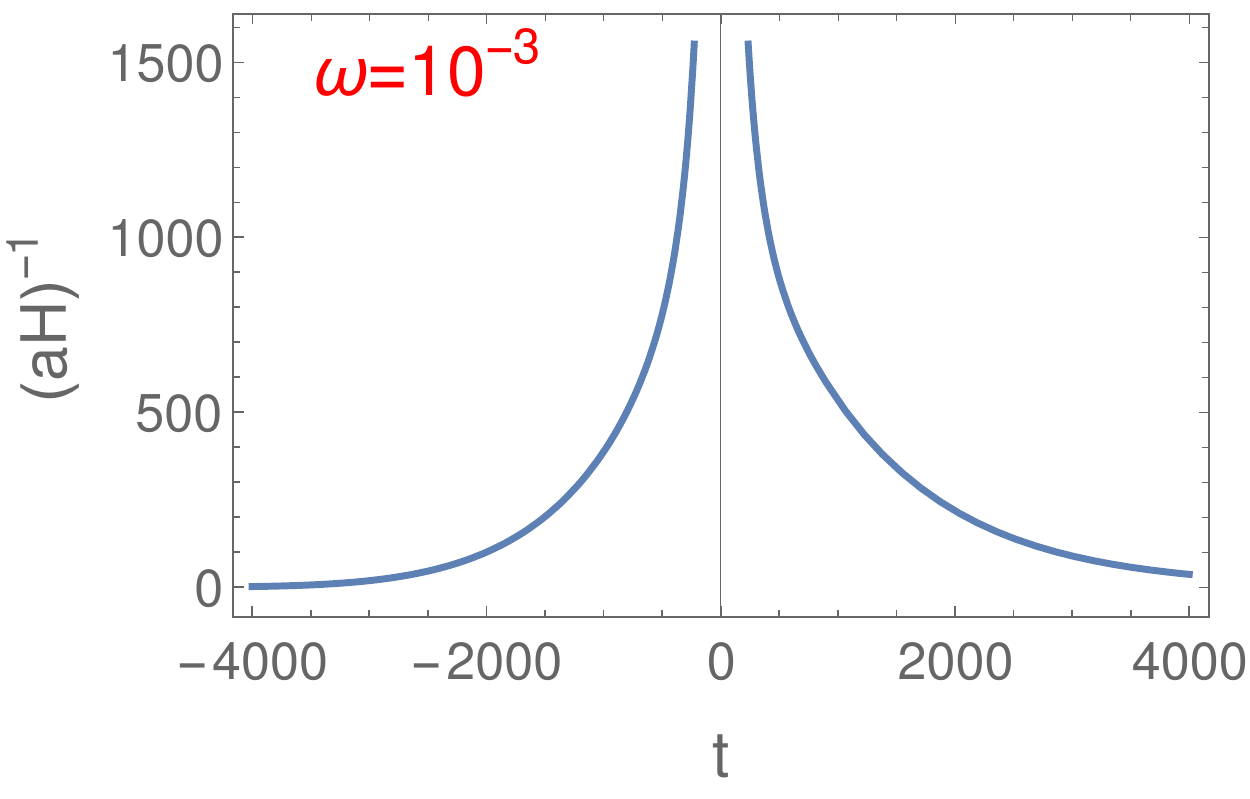}}~~
\subfloat[\label{Fig_newb}]{\includegraphics[scale=0.66]{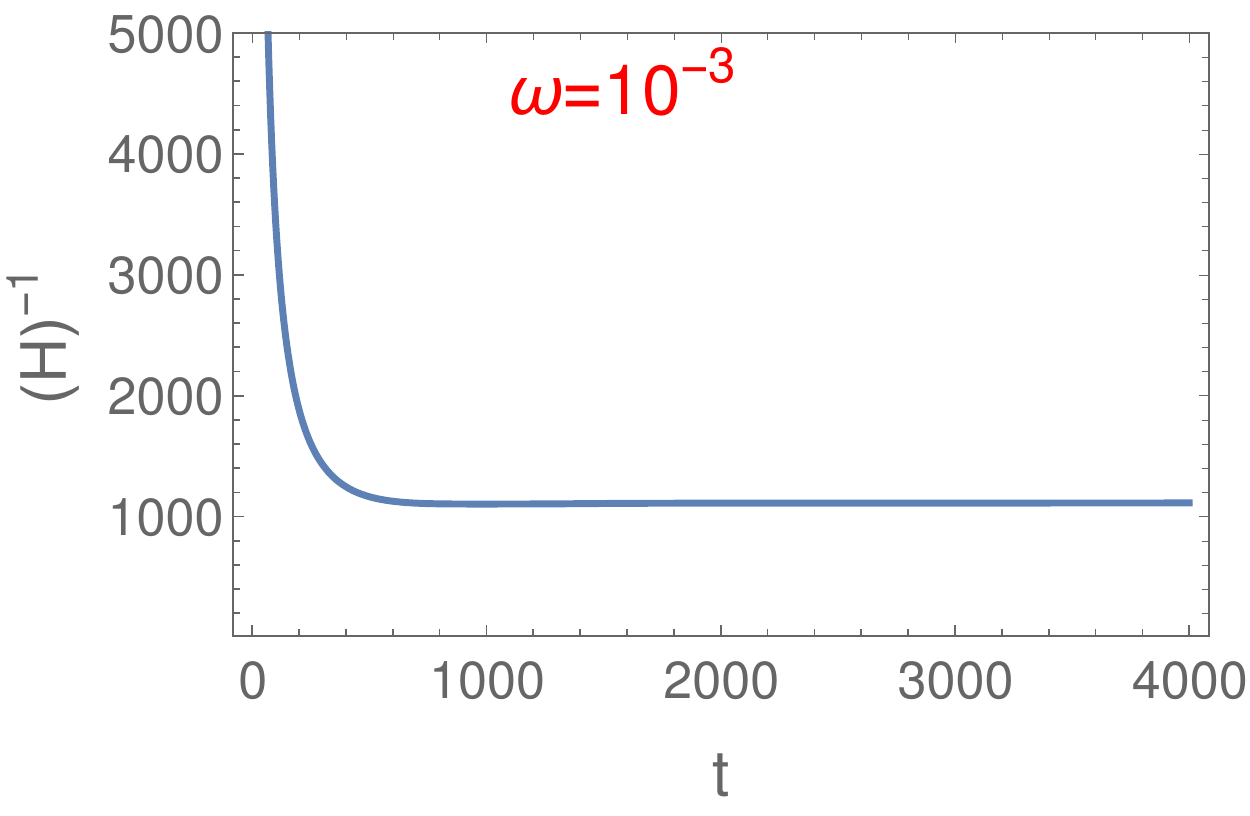}}
\caption{The above figure depicts the time evolution of (a) the comoving Hubble radius $\frac{1}{aH}$ and 
(b) the inverse Hubble parameter $H^{-1}$. Both the above figures are illustrated for $\omega=10^{-3}$.}
\label{Fig_new}
\end{figure*}

Based on the above arguments, before moving to the perturbation calculations, we would like to investigate the asymptotic behaviour of the 
comoving Hubble radius (defined by $\frac{1}{aH}$) in the context of present model. Using the background solution of the Hubble parameter from 
\ref{Fig_3a}, we give the evolution of $\frac{1}{aH}$ with respect to cosmic time in \ref{Fig_newa} which clearly demonstrates that the comoving Hubble 
radius monotonically decreases with time and goes to zero asymptotically on both sides of the bounce. Here it may be mentioned that unlike to 
the comoving Hubble radius, the inverse Hubble parameter does not go to zero asymptotically but reaches to a constant value 
at late stage of the universe, which is depicted in \ref{Fig_newb} showing the behaviour of $H^{-1}$ vs. $t$. 
This corresponds to a late time accelerating universe. 
The asymptotic evolution of the comoving Hubble radius leads to the perturbation modes generate near the 
bouncing regime where the Hubble radius has an infinite size such that all the perturbation modes are contained inside the horizon. 
In this regard the present scenario is different from 
the usual matter bounce models where the Hubble radius diverges asymptotically and the perturbation modes generate far away from the bounce. 
Therefore in the next section we solve the perturbation equations near the bouncing point $t = 0$.

\subsection{Scalar perturbation}
The scalar metric perturbation over FRW metric can be written in the longitudinal gauge as,
\begin{eqnarray}
 ds^2 = a^2(\eta)\bigg[\big(1 + 2\Psi\big)d\eta^2 - \big(1 - 2\Psi\big)\delta_{ij}dx^idx^j\bigg]
 \label{sp1}
\end{eqnarray}
where the line element is expressed in ($\eta$, $\vec{x}$) coordinates with $\eta$ being the conformal time defined by $d\eta = \frac{dt}{a(t)}$ and 
the variable $\Psi(\eta, \vec{x})$ symbolizes the scalar metric fluctuation. Here it may be mentioned that the spacelike and the timelike components of 
scalar perturbation are considered to be same, this is because the background evolution has no anisotropic stress in the present context. Moreover, 
we expand the radion field as,
\begin{eqnarray}
 \Phi(\eta,\vec{x}) = \Phi_0(\eta) + \delta\Phi(\eta,\vec{x})
 \label{sp2}
\end{eqnarray}
in terms of the background radion $\Phi_0(\eta)$ and the fluctuation $\delta\Phi(\eta,\vec{x})$. As a result, the scalar perturbation, in the longitudinal 
gauge, follows the following equations (upto first order perturbations) \cite{Brandenberger:2003vk},
\begin{eqnarray}
\nabla^2\Psi - 3\mathcal{H}\Psi' - 3\mathcal{H}\Psi&=&\frac{\kappa^2}{2}a^2\delta T^0_0\nonumber\\
\big(\Psi' + \mathcal{H}\Psi\big)_{,i}&=&\frac{\kappa^2}{2}a^2\delta T^0_i\nonumber\\
\bigg[\Psi'' + 3\mathcal{H}\Psi' + \big(2\mathcal{H}' + \mathcal{H}^2\big)\Psi\bigg]\delta^i_j&=&-\frac{\kappa^2}{2}a^2\delta T^i_j
\label{sp3}
\end{eqnarray}
where prime denotes $\frac{d}{d\eta}$ and $\mathcal{H} = \frac{a'}{a}$ is the Hubble parameter in conformal time coordinate. The variation of 
matter energy-momentum tensor, i.e, $\delta T_{\mu\nu}$ (recall, the radion field is the only matter field in the present context) present in the 
right hand side of the above equations, can be obtained from \ref{Eq24} and are given by,
\begin{eqnarray}
 \delta T^0_0&=&\frac{1}{a^2}\bigg[G(\Phi_0)\Phi_0'\delta\Phi' + \frac{1}{2}G'(\Phi_0)(\Phi_0')^2\delta\Phi + 2a^2M^3k_0V'(\Phi_0)\delta\Phi\bigg]\nonumber\\
 \delta T^0_i&=&\frac{1}{a^2}\partial_{i}\bigg[G(\Phi_0)\Phi_0'\delta\Phi\bigg]\nonumber\\
 \delta T^{i}_{j}&=&-\frac{1}{a^2}\delta^i_j\bigg[G(\Phi_0)\Phi_0'\delta\Phi' + \frac{1}{2}G'(\Phi_0)(\Phi_0')^2\delta\Phi - 2a^2M^3k_0V'(\Phi_0)\delta\Phi\bigg]
 \label{sp4}
\end{eqnarray}
where we explicitly used the fluctuation of the radion field shown in \ref{sp2} and recall, $V(\Phi)$ and $G(\Phi)$ are the radion potential and 
the non-canonical kinetic term of the radion field respectively. In \ref{sp4} and in the rest of the discussion the primes in $V(\Phi_0)$ and $G(\Phi_0)$ are with respect to the background radion field $\Phi_0$ while the primes in $\mathcal{H}$ and $\Phi_0$ are with respect to the conformal time $\eta$. Substituting the expressions of $\delta T_{\mu\nu}$ into the
set of equations \ref{sp3}, we get
\begin{eqnarray}
 \nabla^2\Psi - 3\mathcal{H}\Psi' - 3\mathcal{H}\Psi&=&\frac{\kappa^2}{2}
 \bigg[G(\Phi_0)\Phi_0'\delta\Phi' + \frac{1}{2}G'(\Phi_0)(\Phi_0')^2\delta\Phi + 2a^2M^3k_0V'(\Phi_0)\delta\Phi\bigg]\nonumber\\
 \Psi' + \mathcal{H}\Psi&=&\frac{\kappa^2}{2}\Phi_0'\delta\Phi\nonumber\\
 \Psi'' + 3\mathcal{H}\Psi' + \big(2\mathcal{H}' + \mathcal{H}^2\big)\Psi&=&\frac{\kappa^2}{2}
 \bigg[G(\Phi_0)\Phi_0'\delta\Phi' + \frac{1}{2}G'(\Phi_0)(\Phi_0')^2\delta\Phi - 2a^2M^3k_0V'(\Phi_0)\delta\Phi\bigg]
 \label{sp5}
\end{eqnarray}
respectively. The second of \ref{sp5} can be used to obtain $\delta\Phi$ in terms of $\Psi$ and $\Psi'$, substituting which into the other two 
equations leads to the evolution of $\Psi(\eta,\vec{x})$ as,  
\begin{eqnarray}
 \Psi'' - \nabla^2\Psi + 6\mathcal{H}\Psi' + \big(2\mathcal{H}' + 4\mathcal{H}^2\big)\Psi = 
 -4a^2M^3k_0\bigg(\frac{V'(\Phi_0)\big(\Psi' + \mathcal{H}\Psi\big)}{G(\Phi_0)\Phi_0'}\bigg)
 \label{sp6}
\end{eqnarray}
which explicitly depends on the non-canonical term $G(\Phi_0)$ and for $G(\Phi_0) = 1$, \ref{sp6} reduces to that of the canonical scalar field case 
\cite{Brandenberger:2003vk}. 
To solve the above perturbation equation, we will use the background evolution of the Hubble parameter and the radion field, which are obtained 
in the cosmic time coordinate. Thus we first transform \ref{sp6} in terms of the cosmic time and for this purpose, we need the following relations,
\begin{eqnarray}
 \Psi' = a\dot{\Psi}~~~~~~~~~~~~~~~~~~~~~\rm{and}~~~~~~~~~~~~~~~~~~~~\Psi'' = a^2\ddot{\Psi} + a^2H\dot{\Psi}
 \nonumber
\end{eqnarray}
with overdot and prime representing $\frac{d}{dt}$ and $\frac{d}{d\eta}$ respectively. As a result, \ref{sp6} turns out to be,
\begin{eqnarray}
 \ddot{\Psi} - \frac{1}{a^2}\nabla^2\Psi + \bigg[7H + \frac{2k_0^2~V'(\xi_0)}{3c_2^2G(\xi_0)\dot{\xi}_0}\bigg]\dot{\Psi} 
 + \bigg[2\dot{H} + 6H^2 + \frac{2k_0^2H~V'(\xi_0)}{3c_2^2G(\xi_0)\dot{\xi}_0}\bigg]\Psi = 0
 \label{sp8}
\end{eqnarray}
where $H = \frac{\dot{a}}{a}$ is the Hubble parameter in cosmic time and recall, $\xi_0 = \frac{\Phi_0}{f}$ (with $f = \frac{\sqrt{6M^3c_2^2}}{k_0}$) 
is the dimensionless radion field. \ref{sp8} clearly reveals how the dynamics of the scalar perturbation (i.e the acceleration 
term $\ddot{\Psi}$) depends on the background evolution of $H(t)$ and $\xi_0(t)$. In particular, the second term in the left hand side leads to an 
oscillation of $\Psi$, the third term denotes a friction term and the fourth term indicates a restoring force. 
As mentioned earlier, the perturbation modes generate near the bounce and thus we are interested to solve the perturbation 
equations near $t = 0$, in which case, the background Hubble parameter and the radion field evolution follow \ref{Eq46}. Using such 
background evolution of $H(t)$ and $\xi(t)$ along with the near-bounce expression of $G(\xi)$ (see \ref{Eq34}), 
we determine $\frac{V'(\xi_0)}{G(\xi_0)\dot{\xi}_0}$ (present in the above equation) as,
\begin{eqnarray}
 \frac{V'(\xi_0)}{G(\xi_0)\dot{\xi}_0}
 &=&\frac{36\omega^2\delta^2(t)}{\dot{\delta}\big[\ln{\frac{\omega}{c_2}} + 2\big(2 - \ln{\frac{\omega}{c_2}}\big)\delta(t)\big]}\nonumber\\
 &=&-\frac{48B\omega c_2\sinh^2(B\pi/8)}{k_0\big(3 - 2e^{B\pi/4}\big)\big(2 - \ln{\frac{\omega}{c_2}}\big)} 
 + \frac{72\omega^2\big(2 - 2\cosh(B\pi/4) + \sinh(B\pi/4)\big)}{\big(3 - 2e^{B\pi/4}\big)^2\big(2 - \ln{\frac{\omega}{c_2}}\big)}~t
 \label{sp9}
\end{eqnarray}
where $B = \frac{A}{6}\frac{\omega}{c_2}\sqrt{\frac{3}{\ln{\big(\frac{c_2}{\omega}}\big)}}$ and 
we retain the expression of $\frac{V'(\xi_0)}{G(\xi_0)\dot{\xi}_0}$ up to the leading order in $t$. With the above expression, 
\ref{sp8} turns out to be,
\begin{eqnarray}
 \ddot{\Psi} - \nabla^2\Psi + \big[-\sqrt{\alpha}p + (q + 14)\alpha t\big]\dot{\Psi} + \big[4\alpha - 2\alpha\sqrt{\alpha}p~t\big]\Psi(\vec{x},t) = 0
 \label{sp10}
\end{eqnarray}
near the bounce (i.e in the leading order of $t$), where $\alpha=\frac{6k_0^2\omega^2}{c_2^2}$ and $p$ and $q$ have the following expressions,
\begin{eqnarray}
 p = 16\sqrt{\frac{2}{3}}\bigg(\frac{B\sinh^2(B\pi/8)}{\big(3 - 2e^{B\pi/4}\big)\big(2 - \ln{\frac{\omega}{c_2}}\big)}\bigg)~~~~~\mathrm{and}~~~~~
 q = \frac{8\big(2 - 2\cosh(B\pi/4) + \sinh(B\pi/4)\big)}{\big(3 - 2e^{B\pi/4}\big)^2\big(2 - \ln{\frac{\omega}{c_2}}\big)}
 \nonumber
\end{eqnarray}
respectively. In terms of the Fourier transformed scalar perturbation variable 
$\Psi_k(t) = \int d\vec{x}e^{-i\vec{k}.\vec{x}}\Psi(\vec{x},t)$, \ref{sp10} can be written as,
\begin{eqnarray}
 \ddot{\Psi}_k + \big[-\sqrt{\alpha}p + (q + 14)\alpha t\big]\dot{\Psi}_k + \big[k^2 + 4\alpha - 2\alpha\sqrt{\alpha}p~t\big]\Psi_k(t) = 0
 \label{sp11}
\end{eqnarray}
Solving \ref{sp11} for $\Psi_k(t)$, we get
\begin{eqnarray}
 \Psi_k(t) = b_1(k)~exp\bigg[\sqrt{\alpha}pt - 7\alpha t^2 - \frac{q}{2}\alpha t^2\bigg]~H\bigg[-1 + \frac{k^2 + 4\alpha}{\alpha(q+14)}, 
 \frac{-p + (q+14)\sqrt{\alpha}~t}{\sqrt{2(q+14)}}\bigg]
 \label{sp12}
\end{eqnarray}
with $H[n,x]$ is the n-th order Hermite polynomial. $b_1(k)$ is the integration constant which can be determined from the initial Bunch-Davies vacuum 
condition given by $\lim_{\eta\rightarrow0}v_k(\eta) = \frac{1}{\sqrt{2k}}e^{-ik\eta}$, where $v_k(\eta)$ is the canonical Mukhanov-Sasaki variable. 
The Bunch-Davies vacuum choice is justified since the primordial modes at $t = 0$ (or equivalently $\eta = 0$) are well 
inside the Hubble horizon. The Bunch-Davies vacuum condition on the Mukhanov-Sasaki variable immediately leads the corresponding condition on 
$\Psi_k(t)$ from the following relation \cite{Brandenberger:2003vk},
\begin{eqnarray}
 \lim_{t \rightarrow 0}\Psi_k(t) = \frac{\kappa^2f}{2k^2}~\lim_{t \rightarrow 0}\big[\sqrt{G(\xi)}~\dot{\xi}v_k'(\eta)\big] 
 = \frac{i\kappa^2f}{2\sqrt{2}k^{3/2}}~\lim_{t \rightarrow 0}\big[\sqrt{G(\xi)}~\dot{\xi}\big]
 \label{sp13}
\end{eqnarray}
and by using the background evolution of $\xi(t)$ along with the expression of $G(\xi)$, we determine the initial condition of 
$\Psi_k(t)$ as follows,
\begin{eqnarray}
 \lim_{t \rightarrow 0}\Psi_k(t) = \frac{\sqrt{3}}{2k^{3/2}}\bigg(\frac{\omega}{c_2}\bigg)\bigg(\frac{k_0}{M}\bigg)^{3/2}
 e^{B\pi/4}\big(3 - 2e^{B\pi/4}\big)^{1/2}
 \label{sp14}
\end{eqnarray}
This makes the integration constant $b_1(k)$ have the following form,
\begin{eqnarray}
 b_1(k) = \frac{\sqrt{3}}{2k^{3/2}}\bigg(\frac{\omega}{c_2}\bigg)\bigg(\frac{k_0}{M}\bigg)^{3/2}~\bigg\{\frac{e^{B\pi/4}\big(3 - 2e^{B\pi/4}\big)^{1/2}}
 {H\big[-1 + \frac{k^2 + 4\alpha}{\alpha(q+14)}, \frac{-p}{\sqrt{2(q+14)}}\big]}\bigg\}
 \nonumber
\end{eqnarray}
Substituting the above expression of $b_1(k)$ into \ref{sp12} yields the following solution for the scalar perturbation variable,
\begin{eqnarray}
 \Psi_k(t) = \frac{\sqrt{3}}{2k^{3/2}}\bigg(\frac{\omega}{c_2}\bigg)\bigg(\frac{k_0}{M}\bigg)^{3/2}e^{B\pi/4}\big(3 - 2e^{B\pi/4}\big)^{1/2} 
 e^{[p\sqrt{\alpha}~t~ - 7\alpha t^2 - \frac{q}{2}\alpha t^2]}~\Bigg\{\frac{H\big[-1 + \frac{k^2 + 4\alpha}{\alpha(q+14)}, 
 \frac{-p + (q+14)\sqrt{\alpha}~t}{\sqrt{2(q+14)}}\big]}{H\big[-1 + \frac{k^2 + 4\alpha}{\alpha(q+14)}, \frac{-p}{\sqrt{2(q+14)}}\big]}\Bigg\}
 \label{sp15}
\end{eqnarray}
where $p$ and $q$ are given below \ref{sp10}. Consequently the solution of $\Psi_k(t)$ immediately leads to the scalar power spectrum for 
$k$-th modes as,
\begin{eqnarray}
 P_{\Psi}(k,t)&=&\frac{k^3}{2\pi^2}\bigg|\Psi_k(t)\bigg|^2\nonumber\\
  &=&\frac{3}{8\pi^2}\bigg(\frac{\omega}{c_2}\bigg)^2\bigg(\frac{k_0}{M}\bigg)^{3}e^{B\pi/2}\big(3 - 2e^{B\pi/4}\big) 
 e^{[2p\sqrt{\alpha}~t~ - 14\alpha t^2 - q\alpha t^2]}~\Bigg\{\frac{H\big[-1 + \frac{k^2 + 4\alpha}{\alpha(q+14)}, 
 \frac{-p + (q+14)\sqrt{\alpha}~t}{\sqrt{2(q+14)}}\big]}{H\big[-1 + \frac{k^2 + 4\alpha}{\alpha(q+14)}, \frac{-p}{\sqrt{2(q+14)}}\big]}\Bigg\}^2\nonumber\\
 \label{sp16}
\end{eqnarray}
Here we would like to mention that our main aim in this section is to investigate whether the theoretical predictions of 
$n_s$, $\mathcal{A}_s$ and $r$ match with the Planck 2018 results which put a constraint on these observable quantities 
around the CMB scale. Therefore the scale of interest in the present context is around the CMB scale given by 
$k_{CMB} \approx 0.02\mathrm{Mpc}^{-1} \approx 10^{-40}\mathrm{GeV}$. With the background solution of Hubble parameter from 
\ref{Eq46}, we determine the expression of the time when $k_{CMB}$ crosses the horizon by using the horizon crossing 
relation $k = aH$, and is given by,
\begin{eqnarray}
 t_{h} = \frac{k_{CMB}}{12k_0^2}\bigg(\frac{c_2^2}{\omega^2}\bigg)~~,
 \label{new1}
\end{eqnarray}
where, $t_h$ is the horizon crossing time of the CMB scale and recall, $k_0$ being the bulk curvature scale. 
As we will show later that the model stands to be a viable one in regard to the Planck constraints for the parameter ranges :
$\omega = 10^{-3}$  and $\frac{k_0}{M} = [0.601 , 0.607]$ respectively. Such parametric ranges make the horizon crossing instance of $k_{CMB}$ 
as $t_{h} \sim 10^{-68}\mathrm{GeV}^{-1} \approx 10^{-93}\mathrm{sec}$ (the conversion $1\mathrm{GeV}^{-1} = 10^{-25}\mathrm{sec}$ may be useful). 
This estimation of $t_{h}$ along with \ref{Eq46} indicate that the scale factor, around the horizon crossing instance of $k_{CMB}$, 
practically behaves as $a(t \simeq t_{h}) = 1 + 6k_0^2\big(\frac{\omega^2}{c_2^2}\big)t^2$; which in turn confirms the fact 
that the CMB scale crosses the horizon near the bouncing regime. Correspondingly, the scalar power spectrum at horizon crossing 
can be expressed as,
\begin{eqnarray}
 P_{\Psi}(k,t)\bigg|_{h.c} = \frac{3}{8\pi^2}\bigg(\frac{\omega}{c_2}\bigg)^2\bigg(\frac{k_0}{M}\bigg)^3e^{B\pi/2}\big(3 - 2e^{B\pi/4}\big) 
 e^{[2p\sqrt{\alpha}~t_h - 14\alpha t_h^2 - q\alpha t_h^2]}~\bigg\{\frac{H\big[-1 + \frac{k^2 + 4\alpha}{\alpha(q+14)}, 
 \frac{-p + (q+14)\sqrt{\alpha}~t_h}{\sqrt{2(q+14)}}\big]}{H\big[-1 + \frac{k^2 + 4\alpha}{\alpha(q+14)}, \frac{-p}{\sqrt{2(q+14)}}\big]}\bigg\}^2~~~.
 \label{sp17}
\end{eqnarray}
With \ref{sp17}, we can determine the observable quantities like the scalar 
spectral index of the primordial curvature
perturbations ($n_s$), the scalar perturbation amplitude ($A_s$) etc. 
However before proceeding to calculate $n_s$ and $A_s$, we will perform first the tensor perturbation,
which is necessary for evaluating the tensor-to-scalar ratio ($r$).

\subsection{Tensor perturbation}
In this section we consider the tensor
perturbation on the FRW metric background which is defined as follows,
\begin{eqnarray}
 ds^2 = -dt^2 + a(t)^2\left(\delta_{ij} + h_{ij}\right)dx^idx^j\, ,
 \label{ten per metric}
\end{eqnarray}
where $h_{ij}(t,\vec{x})$ is the tensor perturbation. The variable $h_{ij}(t,\vec{x})$
is itself a gauge invariant quantity, and the tensor perturbed action up to quadratic order is given by \cite{Hwang:2005hb,Noh:2001ia,Hwang:2002fp},
\begin{eqnarray}
 \delta S_{h} = \int dt d^3\vec{x} a(t) z_T(t)^2\left[\dot{h}_{ij}\dot{h}^{ij}
 - \frac{1}{a^2}\left(\partial_lh_{ij}\right)^2\right]\, ,
 \label{ten per action}
\end{eqnarray}
where $z_T(t)$, in the non-canonical scalar-tensor theory i.e the case of the present context, has the following form \cite{Hwang:2005hb},
\begin{eqnarray}
 z_T(t) = \frac{a(t)}{\kappa}\, ,
 \label{ten per z}
\end{eqnarray}
\ref{ten per action} indicates that the speed of the tensor perturbation is $c_T^2 = 1$ i.e
the gravitational waves propagate with the speed of light which is
unity in the natural units. This is in agreement with the event GW170817 according to which, the gravitational wave and the electromagnetic wave have 
the same propagation speed. At this stage, it deserves mentioning that the speed of the gravitational wave depends on the background model, as for example, 
the $c_T^2$ is not unity in scalar-Einstein-Gauss-Bonnet (GB) gravity theory and the deviation of $c_T^2$ from unity is proportional to the GB coupling 
function considered in the model. However there exists a certain class of GB coupling function for which the gravitational wave propagates 
with $c_T^2 = 1$ leading to the compatibility of the GB model with GW170817 (the bouncing phenomenology in such a class of Gauss-Bonnet theory which is compatible 
with GW170817 has been recently discussed in \cite{Elizalde:2020zcb}). On other hand, 
the non-canonical scalar-tensor theory always leads to $c_T^2 = 1$ irrespective of the form of the non-canonical coupling function. 
Coming back to \ref{ten per z}, the tensor perturbation is ensured to be
stable in the present context as the condition $z_T(t)^2 = \frac{a(t)^2}{\kappa^2} > 0$ holds. The action 
\ref{ten per action} leads to the following equation for the tensor
perturbed variable $h_{ij}$,
\begin{eqnarray}
  \frac{1}{a(t)z_T^2(t)}\frac{d}{dt}\bigg[a(t)z_T^2(t)\dot{h}_{ij}\bigg] - \frac{1}{a^2}\partial_{l}\partial^{l}h_{ij} = 0
  \label{ten per eom1}
 \end{eqnarray}
The Fourier transformed tensor perturbation variable is defined as $h_{ij}(t,\vec{x}) = \int d\vec{k}~\sum_{\gamma}\epsilon_{ij}^{(\gamma)}~
h_{(\gamma)}(\vec{k},t) e^{i\vec{k}.\vec{x}}$, where $\gamma = '+'$ and $\gamma = '\times'$ represent two polarization modes. Moreover
$\epsilon_{ij}^{(\gamma)}$ are the polarization tensors satisfying $\epsilon_{ii}^{(\gamma)} = k^{i}\epsilon_{ij}^{(\gamma)} = 0$. In
terms of the Fourier transformed tensor variable $h_k(t)$, \ref{ten per eom1} can be expressed as,
 \begin{eqnarray}
  \frac{1}{a(t)z_T^2(t)}\frac{d}{dt}\bigg[a(t)z_T^2(t)\dot{h}_k\bigg] + \frac{k^2}{a^2}h_k(t) = 0
  \label{ten per eom2}
 \end{eqnarray}
 The two polarization modes obey the same \ref{ten per eom2} and thus we omit the polarization index. Moreover, both the polarization modes even follow the same initial condition and hence have the same solution. Therefore, in the expression of the tensor power spectrum, we will introduce a multiplicative factor $2$ due to the contribution from both the polarization modes. As mentioned earlier, the perturbation modes generate near the bouncing regime (because at that time all the perturbation modes lie within the Hubble horizon) where the background Hubble parameter ($H(t)$) follow the evolution presented in \ref{Eq46}. From the solution of $H(t)$ the form of the scale factor turns out to be $a(t) = \big(\cosh\big[\frac{6\omega}{c_2}k_0t\big]\big)^{1/3}$ which can be expanded in a Taylor series about $t = 0$ (i.e about the bounce point) as,
 \begin{eqnarray}
  a(t) \simeq 1 + \frac{6\omega^2}{c_2^2}k_0^2t^2 + \mathcal{O}(t^3)
  \nonumber
 \end{eqnarray}
We are interested to solve the perturbation near the bounce (i.e., $t=0$) where the scale factor can be approximated to be
 $a(t) \simeq 1 + \frac{6\omega^2}{c_2^2}k_0^2t^2$. Using this expression of the near-bounce scale factor, we determine $a(t)z_T^2(t)$ as,
 \begin{eqnarray}
  a(t)z_T^2(t) = \frac{a^3(t)}{\kappa^2} \simeq \frac{1}{\kappa^2}\big(1 + 3\alpha t^2\big)
  \label{ten per z2}
 \end{eqnarray}
with $\alpha = 6k_0^2\frac{\omega^2}{c_2^2}$. Substituting this
expression of $a(t)z_T^2(t)$ into \ref{ten per eom2} and
after some algebra, we get the following equation for the Fourier transformed tensor peturbation variable,
\begin{eqnarray}
 \ddot{h}_k + 6\alpha \dot{h}_k~t + k^2h_k(t) = 0
 \label{ten per eom3}
\end{eqnarray}
at leading order in $t$ (since
the perturbation modes generate near the bouncing phase i.e., near $t=0$). Solving \ref{ten per eom3} for $h_k(t)$ ,
we get,
\begin{eqnarray}
 h_k(t) = b_2(k)~e^{-3\alpha t^2}~H\bigg[-1 + \frac{k^2}{6\alpha}, \sqrt{3\alpha}~t\bigg]
 \label{ten per sol1}
\end{eqnarray}
where $b_2(k)$ is an integration constant and can be determined from an initial condition. As an initial condition, we consider that
the tensor perturbation field starts from the adiabatic vacuum, more precisely the initial configuration is given by,
$\lim_{t\rightarrow 0}\big[z_T(t)h_k(t)\big] = \frac{1}{\sqrt{2k}}$. This immediately leads to the expression of $b_2(k)$ as,
\begin{eqnarray}
 b_2(k) = \frac{1}{z_T(t \rightarrow 0)}\bigg[\frac{2\Gamma\big(1 - \frac{k^2}{12\alpha}\big)}{\sqrt{2\pi k}~2^{\frac{k^2}{6\alpha}}}\bigg]
 = \kappa\bigg[\frac{2\Gamma\big(1 - \frac{k^2}{12\alpha}\big)}{\sqrt{2\pi k}~2^{\frac{k^2}{6\alpha}}}\bigg]~~~.
 \label{ten per bc}
\end{eqnarray}
In the second equality of the above equation, we use $z_T(t\rightarrow 0) = 1/\kappa$ from \ref{ten per z}. Putting
this expression of $b_2(k)$ into \ref{ten per sol1} yields the final solution of $h_k(t)$ as follows,
\begin{eqnarray}
 h_k(t) = \bigg(\frac{2\kappa~\Gamma\big(1 - \frac{k^2}{12\alpha}\big)}{\sqrt{2\pi k}~2^{\frac{k^2}{6\alpha}}}\bigg)~
 e^{-3\alpha t^2}~H\bigg[-1 + \frac{k^2}{6\alpha}, \sqrt{3\alpha}~t\bigg]
 \label{ten per sol2}
\end{eqnarray}
\ref{ten per sol2} represents the solution of the tensor perturbation for both the polarization modes. The solution of $h_k(t)$ immediately
leads to the tensor power spectrum as,
\begin{eqnarray}
 P_{h}(k,t)&=&\frac{k^3}{2\pi^2}~\sum_{\gamma}\bigg|h_k^{(\gamma)}(t)\bigg|^2 \nonumber\\
 &=&\frac{2k^2}{\pi^3}~\frac{\bigg(\kappa~\Gamma\big(1 - \frac{k^2}{12\alpha}\big)\bigg)^2}
 {~2^{\frac{k^2}{3\alpha}}} e^{-6\alpha t^2}~
 \bigg\{H\bigg[-1 + \frac{k^2}{6\alpha}, \sqrt{3\alpha}~t\bigg]\bigg\}^2
 \label{ten power spectrum}
\end{eqnarray}
It may be noticed that $\gamma = '+'$ and $\gamma = '\times'$ modes contribute equally to the power spectrum, as expected because their
solutions behave similarly. At the horizon crossing $k=aH\simeq 2\alpha t_h$, the tensor power spectrum turns out to be,
\begin{eqnarray}
 P_{h}(k,t)\bigg|_{h.c} = \frac{12k_0^3\omega^2}{\pi^3M^3c_2^2}~\alpha t_h^2~\frac{\bigg(\kappa~\Gamma\big(1 - \frac{\alpha t_h^2}{3}\big)\bigg)^2}
 {~2^{\frac{4\alpha t_h^2}{3}}} e^{-6\alpha t_h^2}~
 \bigg\{H\bigg[-1 + \frac{2}{3}\alpha t_h^2, \sqrt{3\alpha}~t_h\bigg]\bigg\}^2
 \label{ten power spectrum_H.C}
\end{eqnarray}
with $t_h$ being the horizon crossing instance.

\begin{figure}[!h]
\begin{center}
 \centering
 \includegraphics[scale=1.00]{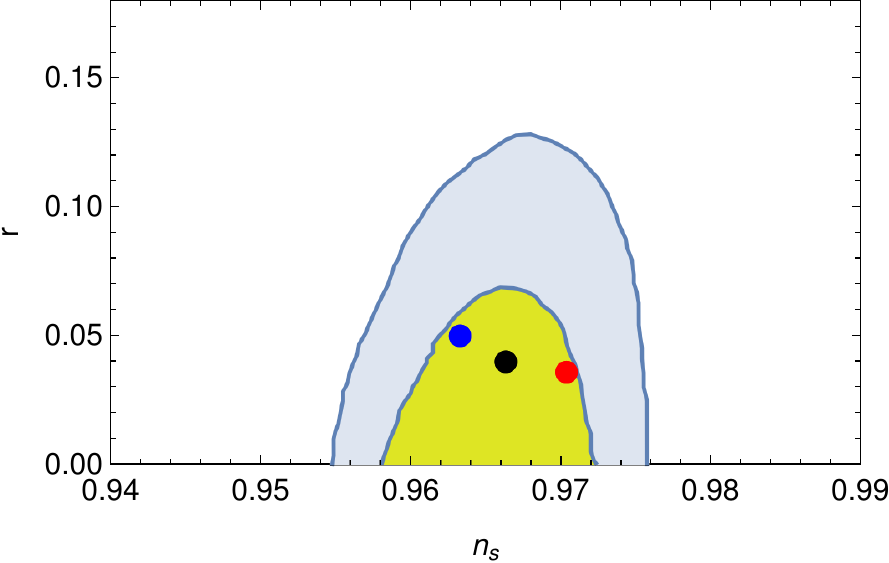}
 \caption{$1\sigma$ (yellow) and $2\sigma$ (light blue) contours for Planck 
 2018 results \cite{Akrami:2018odb}, on $n_s-r$ plane. 
 Additionally, we present the predictions of the present bounce scenario with $\frac{R_h}{\alpha} = 14$ (blue point), $\frac{R_h}{\alpha} = 16$ 
 (black point) and $\frac{R_h}{\alpha} = 19$ (red point).}
 \label{plot_observable}
\end{center}
\end{figure}

Now we can explicitly confront the model at hand with the latest
Planck observational data \cite{Akrami:2018odb}, so we shall
calculate the spectral index of the primordial curvature
perturbations $n_s$ and the tensor-to-scalar ratio $r$, which are defined as follows,
\begin{eqnarray}
 n_s - 1 = \frac{\partial\ln{P_{\Psi}}}{\partial\ln{k}}\bigg|_{H.C}~~~~~~~~,~~~~~~~~~~~
 r = \frac{P_h(k,t)}{P_{\Psi}(k,t)}\bigg|_{H.C}\label{spectral index1}
\end{eqnarray}
As evident from these expressions, $n_s$ and $r$ are evaluated at
the time of the horizon exit near the bouncing point (symbolized by `H.C' in the above equations), for positive times  when
$k=aH$ i.e. when the mode $k$ crosses the Hubble horizon. Using
\ref{sp16}, we determine
$\frac{\partial\ln{P_{\Psi}}}{\partial\ln{k}}$ as follows,
\begin{eqnarray}
 \frac{\partial\ln{P_{\Psi}}}{\partial\ln{k}} = \frac{4k^2}{\alpha(q+14)}
 \bigg\{\frac{H^{(1,0)}\big[-1 + \frac{k^2 + 4\alpha}{\alpha(q+14)}, 
 \frac{-p + (q+14)\sqrt{\alpha}~t}{\sqrt{2(q+14)}}\big]}{H\big[-1 + \frac{k^2 + 4\alpha}{\alpha(q+14)}, 
 \frac{-p + (q+14)\sqrt{\alpha}~t}{\sqrt{2(q+14)}}\big]} - \frac{H^{(1,0)}\big[-1 + \frac{k^2 + 4\alpha}{\alpha(q+14)}, 
 \frac{-p}{\sqrt{2(q+14)}}\big]}{H\big[-1 + \frac{k^2 + 4\alpha}{\alpha(q+14)}, 
 \frac{-p}{\sqrt{2(q+14)}}\big]}\bigg\}
\label{spectral index2}
\end{eqnarray}
where $H^{(1,0)}[z_1,z_2]$ is the derivative of
$H[z_1,z_2]$ with respect to its first argument. 
Therefore \ref{spectral index2} immediately leads to the spectral index as,
\begin{eqnarray}
 n_s = 1 - \frac{4k^2}{\alpha(q+14)}
 \bigg\{\frac{H^{(1,0)}\big[-1 + \frac{k^2 + 4\alpha}{\alpha(q+14)}, 
 \frac{-p}{\sqrt{2(q+14)}}\big]}{H\big[-1 + \frac{k^2 + 4\alpha}{\alpha(q+14)}, 
 \frac{-p}{\sqrt{2(q+14)}}\big]} - \frac{H^{(1,0)}\big[-1 + \frac{k^2 + 4\alpha}{\alpha(q+14)}, 
 \frac{-p + (q+14)\sqrt{\alpha}~t}{\sqrt{2(q+14)}}\big]}{H\big[-1 + \frac{k^2 + 4\alpha}{\alpha(q+14)}, 
 \frac{-p + (q+14)\sqrt{\alpha}~t}{\sqrt{2(q+14)}}\big]}\bigg\}_{h.c}
 \label{spectral index3}
\end{eqnarray}
As mentioned earlier, the perturbation modes are generated and
also cross the horizon near the bounce. Thus we can safely use the
near-bounce scale factor in the horizon crossing condition to
determine $k = aH = 2\alpha t_h$ (where $t_h$ is the horizon
crossing time). Using this relation, \ref{spectral index3}
turns out to be,
\begin{eqnarray}
 n_s = 1 - \frac{16\alpha t_h^2}{(q+14)}
 \bigg\{\frac{H^{(1,0)}\big[-1 + \frac{4(\alpha t_h^2 + 1)}{(q+14)}, 
 \frac{-p}{\sqrt{2(q+14)}}\big]}{H\big[-1 + \frac{4(\alpha t_h^2 + 1)}{(q+14)}, 
 \frac{-p}{\sqrt{2(q+14)}}\big]} - \frac{H^{(1,0)}\big[-1 + \frac{4(\alpha t_h^2 + 1)}{(q+14)}, 
 \frac{-p + (q+14)\sqrt{\alpha}~t_h}{\sqrt{2(q+14)}}\big]}{H\big[-1 + \frac{4(\alpha t_h^2 + 1)}{(q+14)}, 
 \frac{-p + (q+14)\sqrt{\alpha}~t_h}{\sqrt{2(q+14)}}\big]}\bigg\}_{h.c}
 \label{spectral index4}
\end{eqnarray}
Furthermore, the tensor-to-scalar ratio is given by,
\begin{eqnarray}
 r = \frac{P_h(k,t)}{P_\Psi(k,t)}\bigg|_{k=a(t_h)H(t_h)}
 \label{tensor to scalar ratio}
\end{eqnarray}
where the solutions of $P_h$ and $P_\Psi$ are shown in
\ref{ten power spectrum_H.C} and \ref{sp17} respectively.
\ref{spectral index4} and \ref{tensor to scalar ratio}
clearly indicate that $n_s$ and $r$ depend on the dimensionless
parameters $\omega$ and $\alpha t_h^2$ which is further connected to the Ricci
scalar at horizon crossing by $\alpha t_h^2 = \big(\frac{R_h}{12\alpha} - 1\big)$. Therefore, we can argue that
the observable quantities $n_s$ and $r$ depend on $\omega$ and $R_h/\alpha$.
With this information, we now directly confront the theoretical expressions of scalar spectral index \ref{spectral index4}
and tensor-to-scalar ratio \ref{tensor to scalar ratio} derived from the present model with the Planck 2018 constraints
\cite{Akrami:2018odb}. In particular, we estimate the allowed values of $\frac{R_h}{\alpha}$ and $\omega$ which in turn can give rise to $n_s$ and $r$ in agreement with the Planck data. This is presented in \ref{plot_observable} where we compute $n_s$ and $r$ for three choices of $\frac{R_h}{\alpha}$ (viz, $\frac{R_h}{\alpha} = 14$ (blue point), $\frac{R_h}{\alpha} = 16$ (black point) and $\frac{R_h}{\alpha} = 19$ (red point) with $\omega = 10^{-3}$. The allowed values of $r$ and $n_s$ from Planck data within $1-\sigma$ and $2-\sigma$ constraints are illustrated by the yellow and the blue regions respectively in \ref{plot_observable}. We note that with $\omega=10^{-3}$ and all the three aforesaid values of $\frac{R_h}{\alpha}$ the model estimated $n_s$ and $r$ are within the $1-\sigma$ constraints reported by Planck 2018 data.

At this stage it may be mentioned that scalar-tensor models (with single scalar field) 
which exhibit a matter bounce scenario asymptotically, such that the perturbations are generated far away 
from the bouncing point deeply in the contracting regime, are 
generally not consistent with the Planck results since it gives rise to an exactly scale invariant power spectrum \cite{deHaro:2015wda}. 
Such inconsistency with 
Planck observation was also confirmed in \cite{Nojiri:2019lqw} from a slightly 
different viewpoint, namely from an $F(R)$ gravity theory. It turns out that $F(R)$ models can be
equivalently mapped to scalar-tensor ones via conformal transformation of the metric 
and, thus, the inconsistencies of the spectral index in the two different models are well justified. However there exists counter example 
of this argument in the context of two field matter bounce in \cite{Cai:2013kja} where the authors proposed a cosmological evolution which undergoes 
the phases like matter contraction, then a period of ekpyrotic contraction, followed by a non-singular bounce, and then a phase of fast roll expansion. 
In such scenario, it has been showed that the primordial curvature perturbation dominated by a scale invariant component 
while there are other terms which can lead to a scale dependence at small length scales, in particular there is a subdominant $k^{3/2}$ dependence 
in the expression of scalar power spectrum. Unlike to such scenarios, here we demonstrate that a 
scalar-tensor gravity model indeed leads to a viable bouncing model when the primordial perturbations are generated near the bounce.\\
Furthermore the scalar perturbation amplitude ($A_s$) is
constrained to $\ln{\big[10^{10}A_s\big]} = 3.044 \pm 0.014$ 
from the Planck results \cite{Akrami:2018odb}. From \ref{sp16} we note that the amplitude of scalar perturbations $A_s$ not only depends on $\omega$ and $\frac{R_h}{\alpha}$ but also on the ratio of the 5D bulk curvature ($k_0$) and the 5D Planck mass (M) i.e $\frac{k_0}{M}$. In particular, the scalar perturbation 
amplitude becomes $A_s = 9.5\times10^{-9}~\big(\frac{k_0}{M}\big)^3$ when we take $\omega = 10^{-3}$ and $\frac{R_h}{\alpha} = 16$. This is in accordance with the Planck constraints mentioned above provided $\frac{k_0}{M}$ lies within $\frac{k_0}{M} = [0.601 , 0.607]$ such that the bulk curvature is constrained to be less than the 5D Planck mass, which in turn confirms the validity of the background classical solution. 
However, it may be mentioned that the allowed range of $\frac{k_0}{M}$ is sensitive to the choice of $\omega$, i.e. a different $\omega$ will lead to a
different allowed range for the parameter $\frac{k_0}{M}$. As an example, $\omega = 10^{-4}$ leads to the scalar perturbation
amplitude as $A_s = 9.5\times10^{-11}~\big(\frac{k_0}{M}\big)^3$ which becomes consistent with the Planck results 
for $\frac{k_0}{M} > 1$. However with the condition $\frac{k_0}{M} > 1$, the assumption of the background classical solution ceases to hold true, which is not desirable. Therefore, as a whole, the
observable quantities $n_s$, $r$ and $A_s$ are simultaneously
compatible with the Planck constraints for the parameter ranges :
$\omega = 10^{-3}$, $14 \leq \frac{R_h}{\alpha} \leq 19$, $\frac{k_0}{M} = [0.601 , 0.607]$ respectively. Such parametric ranges make the
horizon crossing Ricci scalar of the order $R_h \sim k_0^2\big(\frac{\omega}{c_2}\big)^2 \sim 10^{-8}M^2 = 10^{28}\mathrm{GeV}^2$.\\

Before concluding, we would like to mention that the present paper studies a non-singular bounce 
from a warped braneworld scenario with dynamical branes, which is found to yield a nearly scale-invariant power
spectra of primordial perturbations. However, in the background of the contracting 
era, the anisotropy grows with the scale factor as $a^{-6}$ and thus the contracting stage becomes unstable to 
the growth of anisotropies, which is known as the BKL instability \cite{new1}. Thus similar to many 
other bounce models, except the ekpyrotic bounce scenario \cite{Cai:2013kja,Cai:2013vm,Erickson:2003zm,Garfinkle:2008ei}, 
the present model also suffers from the BKL instability. 
Thereby it may be an interesting study to explore the possible effects
of radion dynamics in an ekpyrotic bounce scenario to avoid the BKL instability. This however may be considered in a future work.

\section{Conclusion}
\label{S5}
We consider a five dimensional warped braneworld scenario with two 3-branes embedded within the 5D spacetime, where the branes have a non-zero 
cosmological constant $\omega$ leading to a non-flat brane geometry. With dynamical branes the interbrane distance is treated as a 
$4-d$ scalar field, the so-called radion or the modulus, which generates its own potential when the $4-d$ effective action is obtained 
as a consequence of compactification of the extra coordinate. Such a radion field is also associated with a non-canonical kinetic term 
at the level of the four dimesional effective action which exhibits a transition from a normal to a phantom regime (i.e 
from $G(\xi) > 0$ to $G(\xi) < 0$) as the radion field goes 
from higher to lower values. With the vanishing of the brane cosmological constant $\omega$, the branes become flat such that 
the radion potential ceases to exist while the radion kinetic term becomes canonical. Such a non-flat warped braneworld scenario 
is important as it can simultaneously address the gauge-hierarchy problem and the stabilization of the modulus without the necessity 
of any additional scalar field of unknown origin. 

The presence of the phantom regime is further interesting as the cosmological evolution of the radion field in the FRW background leads 
to a violation of the null energy condition, necessary to ensure a non-singular bounce in our visible universe. 
This motivates us to explore the prospect of the radion field in triggering a bouncing universe which in turn can potentially avoid 
the Big-Bang singularity. Note that the radion field by which the bounce is driven arises naturally 
from compactification in the effective four-dimensional theory and generates its own potential due to the presence of the brane cosmological constant, 
unlike most of the scalar-tensor bounce models where the scalar potentials are constructed by hand to explain the 
observations and often their origin remains unexplained. 

An analysis of the background cosmological evolution of the Hubble parameter and the radion field reveals that the radion field starts 
its journey from the normal regime (i.e $G(\xi) > 0$ regime) and decreases monotonically in magnitude with cosmic time until it transits 
to the phantom era where the bounce occurs. With further time evolution the radion asymptotically stabilizes to the value $\frac{\omega}{c_2}$ 
which also represents the inflection point of the modulus potential. Such an asymptotic magnitude of the radion field can stabilize the modulus 
to the appropriate value where the gauge-hierarchy issue can also be adequately addressed.  

With the background evolution, we further investigate the cosmological evolution of the scalar and tensor perturbations to the FRW metric 
from the present model. The primordial perturbation modes 
in the present context generate near the bounce because at that
time the relevant perturbation modes are within the horizon, unlike the usual matter bounce scenario where the
perturbation modes generate deeply in the contracting regime far away from the bouncing point. As a result the tensor perturbation is found 
to be suppressed in comparison to the 
scalar perturbation and the ratio of tensor to scalar perturbation amplitude becomes less than unity in accordance with the Planck results. 
Moreover, the speed of propagation of the tensor perturbation $c_T^2$ turns out to be the same as the speed of light, in agreement with the 
event GW170817.
We compute the scalar spectral index $n_s$, the tensor to scalar ratio $r$ and the amplitude of the scalar perturbations $A_s$ from the 
present model which turns out to be pleasantly in agreement with the latest Planck 2018 observations, 
well within the 1-$\sigma$ regime.

\subsection*{Acknowledgments}
This research was partially supported in part by the International Centre for Theoretical Sciences (ICTS) 
for the program - Physics of the Early Universe - An Online Precursor (code: ICTS/peu2020/08).

\end{document}